\documentclass[prb,superscriptaddress,preprint,showpacs,floatfix,12pt]{revtex4}
\usepackage{graphicx}
\usepackage{dcolumn}
\usepackage{bm}
\usepackage{amssymb}
\usepackage{amsmath}

\begin{document}
\bibliographystyle{apsrev}
\title{Lattice Dynamics and Thermal Equation of State of Platinum}
\author{Tao Sun}
\email{tsun@grad.physics.sunysb.edu}
\affiliation{Department of Physics and Astronomy, Stony Brook University,
Stony Brook, New York 11794, USA}
\author{Koichiro Umemoto}
\author{Zhongqing Wu}
\affiliation{Department of Chemical Engineering and Materials Science,
Minnesota Supercomputing Institute, University of Minnesota, Minneapolis,
Minnesota 55455, USA}
\author{Jin-Cheng Zheng}
\affiliation{Condensed Matter Physics \& Materials Science Department,
Brookhaven National Laboratory, Upton, New York 11973, USA}
\author{Renata M. Wentzcovitch}
\affiliation{Department of Chemical Engineering and Materials Science,
Minnesota Supercomputing Institute, University of Minnesota, Minneapolis,
Minnesota 55455, USA}
\date{\today}
\begin{abstract}
Platinum is widely used as a pressure calibration standard. However, the 
established thermal EOS has uncertainties, especially in the high $P$-$T$ 
range. We use density functional theory to calculate the thermal equation of 
state of platinum, up to $550$~GPa and $5000$~K. The static lattice energy is 
computed by using the LAPW method, with LDA, PBE, and the recently proposed 
WC functional. The electronic thermal free energy is evaluated using the 
Mermin functional. The vibrational part is computed within the quasi-harmonic 
approximation using density functional perturbation theory and 
pseudopotentials. Special attention is paid to the influence of the 
electronic temperature to the phonon frequencies. We find that in overall 
LDA results agree best with the experiments. Based on the DFT calculations
and the established experimental data, we develop a consistent thermal 
EOS of platinum as a reference for pressure calibration.
\end{abstract}

%
%
\maketitle

%
%
\section{Introduction}
Platinum~(Pt) is a widely used high-pressure standard.  Its equation of 
state~(EOS) at room temperature has been established by reducing shock 
Hugoniot \cite{mcqueen, morgan, jamieson, holmes} and by {\it ab initio} 
linear-muffin-tin-orbital~(LMTO) calculations\cite{holmes} up to $660$~GPa. 
Mao {\it et al.}\cite{mao} used the EOS developed in 
Ref.~\onlinecite{holmes}~(Holmes {\it et al.}) to calibrate pressure in their 
compression experiment on Fe and Fe-Ni alloy. The bulk moduli measured at 
the earth's core pressure are substantially higher than those extrapolated 
from seismological observations.\cite{stacey, singh} A large pressure offset 
is needed to remove the discrepancy: about $8\%$ at $100$~GPa, $15\%$ at 
$300$~GPa. The origin of this offset is under investigation. One
possibility is the EOS of Holmes {\it et al.} seriously overestimates 
pressure.\cite{stacey} Singh raised other possibilities.\cite{singh}
He noticed that the one-parameter EOS of platinum agrees with 
the EOS of Holmes {\it et al.} to 1~\% at high pressures, and concluded that 
large systematic error in pressure scale is unlikely. He further 
proposed that the discrepancy is due to the pressure on the sample is 
different from the one on the pressure standard, in the high-pressure X-ray 
diffraction measurements.

There are conflicting reports on the uncertainties of Holmes 
{\it et al.}'s EOS. Dewaele {\it et al.}\cite{dewaele} measured the EOS of 
six metals at ambient temperature to $94$ GPa using a diamond anvil cell~(DAC).
By cross-checking different pressure scales they found Holmes 
{\it et al.}'s EOS overestimates pressure by $\approx 4$ GPa near $100$ GPa at 
room temperature. This conclusion is confirmed by other 
groups.\cite{fei2007,zha} While more recent calculations based 
on density functional theory~(DFT) suggest Holmes {\it et al.}'s EOS 
underestimates, rather than overestimates, pressure.
Xiang {\it et al.}\cite{xiang} computed the thermal 
equation of state of platinum using LMTO and a mean field potential method. 
The pressure they obtained is $5$ to $6$~\% higher than that of 
Holmes {\it et al.} at high pressures. 
Men\'{e}ndez-Proupin {\it et al.} \cite{proupin} reached similar conclusion 
using pseudopotentials. Both calculations employ the local density 
approximation~(LDA) functional.\cite{pz} And the excess pressure is 
attributed to LDA in Ref.~\onlinecite{proupin}. However it can also be 
caused by other factors. In Table II of Ref.~\onlinecite{xiang}, the 
equilibrium volume decreases as the temperature increases. The electronic 
thermal pressure is negative according to this calculation, which is contrary
to expectations.  Ref.~\onlinecite{proupin} uses an ultrasoft 
Rappe-Rabe-Kaxiras-Joannopoulos pseudopotential from the PWSCF 
website,\cite{pseudo} which contains only $5d$, $6s$ and $6p$ valence states. 
Its large cutoff radius~($2.6$ a.u.) may cause error in studying the highly 
compressed structure.

Besides the room temperature isotherm, accurate thermal 
pressure~($P_{\rm th}$) is needed to calibrate pressure in simultaneous 
high-pressure and high-temperature experiments. Experiments cannot easily 
determine $P_{\rm th}$ over a wide temperature and volume range.\cite{fei2004} 
Consequently  $P_{\rm th}$ is often estimated by assuming it is linear in 
temperature and independent of volume.\cite{mcqueen, holmes} Theory can in 
principle do better. In quasi-harmonic approximation~(QHA), DFT calculations 
give $P_{\rm th}$ at any particular temperature and volume. It is desirable
to combine the experimental data with DFT calculations, taking the advantages
of both, and construct a more accurate thermal EOS for pressure calibration.

In this paper we have three goals: first is to check the accuracy of the
theoretical EOS of platinum predicted by different exchange correlation
functionals. In contrast with previous calculations, we find the room 
temperature isotherm computed with LDA lies below, and nearly parallel to 
the experimental compression data. The Fermi level of platinum lies in the
$d$ band and gives a very large density of state~(DOS) $N(E_{\rm F})$. Its 
vibrational frequencies are more sensitive to the electronic temperature 
than those of many other metals. A Kohn anomaly has been observed in platinum 
at $90$~K.\cite{dutton} It becomes weaker and finally disappears when the 
temperature increases. Thus our second goal is to discuss the electronic 
temperature dependence of vibrations~(ETDV) and its influence on 
the thermal properties. Our last goal is to provide an accurate thermal 
EOS for pressure calibration. For this purpose we make corrections 
to the raw DFT results. We correct the room temperature Gibbs free energy 
$G(P,300\rm~K)$ to ensure that it reproduces the experimental isotherm, then 
combine it with the DFT calculated temperature dependence to
get $G(P,T)$. The thermal EOS and thermal properties deduced from the 
corrected Gibbs free energy are in good agreement with the known experimental 
data.

%
%
\section{Computational Method}
The EOS of a material is determined by its Helmholtz free energy
$F(V,T)$, which consists of three parts:
\begin{equation}
F(V,T)=U(V)+F_{\rm vib}(V,T)+F_{\rm ele}(V,T),
\label{eq:free_energy}
\end{equation}
where $U(V)$ is the static energy of the lattice, $F_{\rm vib}(V,T)$ is
the vibrational free energy, $F_{\rm ele}(V,T)$ accounts for the thermal
excitation of the electrons. $U(V)$ is calculated by using the 
linearized augmented plane-wave~(LAPW) method\cite{wien} and three different
exchange-correlation functionals: LDA, Perdew-Burke-Ernzerhof~(PBE),\cite{pbe} 
and Wu-Cohen~(WC).\cite{wc} The $4f$, $5s$, $5p$, $5d$, $6s$ are described as 
valence states, others are treated as core states. The convergence 
parameter $RK_{\rm max}$ is $10.0$, and the muffin-tin radius $R$ is 2.08 a.u.. 
A $16\times16\times16$ Monkhorst-Pack\cite{monkhorst} uniform k-grid is used 
and the integration over the whole Brillouin zone is done with the tetrahedron 
method.\cite{tetra} All the calculations using LAPW are performed with and 
without spin-orbit effect. 

In contrast with the static lattice energy $U(V)$, which is sensitive
to the relaxation of the core states and requires a full-potential treatment,
thermal excitations contribute to much smaller energy variations and mostly 
depend on the valence states. We use pseudopotentials to compute the thermal 
effects. An ultra-soft Vanderbilt pseudopotential\cite{vanderbilt} is 
generated from the reference atomic configuration $5s^25p^65d^96s^16p^0$, 
including non-linear core corrections.\cite{louie} There are two projectors 
in the $s$ channel, $5s$ and $6s$; two in the $p$ channel, $5p$ and an 
unbound $p$ at $0.2$ Ry above the vacuum level; one in the $d$ channel, $5d$. 
The local component is set in the $f$ channel at the vacuum level. The cutoff 
radii for each channel $s$, $p$, $d$ and local are 1.8, 1.9, 1.9 and 1.8 a.u., 
respectively. We use the scalar relativistic approximation and spin-orbit 
effect is not included. This pseudopotential reproduces the LAPW electronic 
band structure, both at the most contracted volume and the $0$~GPa experimental 
volume. We find pseudopotentials with different exchange-correlation 
functionals yield very similar electronic band structures for platinum, and 
we use LDA to compute all the thermal effects. 

We consider $20$ different volumes, with lattice constants from 7.8 to 
6.2 a.u.(17.58 \AA$^3$ to 8.83 \AA$^3$ in volume). For each volume $V_i$, 
we use LAPW to compute its static energy $U(V_i)$ and the LDA pseudopotental 
to evaluate its thermal free energy $F_{\rm vib}(V_i,T)$ and 
$F_{\rm ele}(V_i,T)$. $F_{\rm vib}(V_i,T)$ is treated within QHA with phonon
frequencies dependent on electronic temperature~(denoted as eQHA) as
\begin{equation}
F_{\rm vib}^{\rm eQHA}(V_i,T)=\frac{1}{2}\sum_{\mathbf{q},j}\hbar
\omega_{\mathbf{q},j}(V_i,T_{\rm ele})
+k_{\rm B}T\sum_{\mathbf{q},j}\ln(1-\exp(\frac{-\hbar
\omega_{\mathbf{q},j}(V_i,T_{\rm ele})}{k_{B}T})),
\label{eq:fph}
\end{equation}
where $\omega_{\mathbf{q},j}(V_i,T_{\rm ele})$ denotes the phonon frequency
computed at volume $V_i$ and electronic temperature $T_{\rm ele}$. In 
thermal equilibrium the system temperature $T$, the ionic temperature 
$T_{\rm ion}$, and $T_{\rm ele}$ are equal. We distinguish these three 
temperatures to emphasis the temperature dependence of phonon frequencies 
come from different sources. Anharmonic phonon-phonon interactions cause 
phonon frequencies to depend on $T_{\rm ion}$, but they are omitted in QHA. 
Electronic thermal excitations disturb the charge distribution in the 
crystal and cause phonon frequencies depend on $T_{\rm ele}$. In the 
normal QHA used for insulators and some metals, this effect is also ignored
and $\omega_{\mathbf{q},j}$ has no temperature dependence~(except through 
$V(T)$). Platinum has a larger $N(E_{\rm F})$ than many other metals and 
ETDV may have noticeable effects on its thermal properties. To quantitatively 
measure the influence of ETDV, we compare the vibrational free energies at 
volume $V_i$ and temperature $T_j$~($T_j$=$500$, $1000$, \dots $5000$ K) 
computed without/with ETDV. Without ETDV~(normal QHA), phonon frequencies are 
computed at $T_{\rm ele}$=$0$ K by using 
Methfessel-Paxton\cite{methfessel}~(MP) smearing with a smearing 
parameter of $0.01$ Ry. The corresponding vibrational free energy is denoted 
as $F_{\rm vib}^{\rm QHA}(V_i,T_j)$. With ETDV~(eQHA) phonon frequencies have 
to be computed separately for each $T_j$. This is achieved 
by using the Mermin functional\cite{mermin} and Fermi-Dirac~(FD) smearing. 
The corresponding vibrational free energy is denoted as 
$F_{\rm vib}^{\rm eQHA}(V_i,T_j)$. The difference between these two, 
$\Delta F_{\rm ETDV}(V_i,T_j)$=$F_{\rm vib}^{\rm eQHA}(V_i,T_j)-
F_{\rm vib}^{\rm QHA}(V_i,T_j)$, describes the correction caused by ETDV. 
To get $\Delta F_{\rm ETDV}$ at arbitrary temperature between $0$-$5000$ K 
we fit a 4th order polynomial from $\Delta F_{\rm ETDV}(V_i,T_j)$ 
\begin{eqnarray}
\Delta F_{\rm ETDV}(V_i,T)
&=&F_{\rm vib}^{\rm eQHA}(V_i,T)-F_{\rm vib}^{\rm QHA}(V_i,T)\nonumber\\
&=&a_1(V_i)\cdot T+a_2(V_i)\cdot T^2+a_3(V_i)\cdot T^3+a_4(V_i)\cdot T^4.
\label{eq:phdf}
\end{eqnarray}
The final vibrational free energy is computed as 
$F_{\rm vib}(V_i,T)$=$F_{\rm vib}^{\rm QHA}(V_i,T)+
\Delta F_{\rm ETDV}(V_i,T)$~(we omit the subscript `eQHA' 
and denote $F_{\rm vib}^{\rm eQHA}$ as $F_{\rm vib}$).

Phonon frequencies in the above procedure are determined by density 
functional perturbation theory~(DFPT)\cite{baroni} as implemented in the 
Quantum ESPRESSO\cite{pwscf} package. The dynamical matrices are computed on 
an $8\times8\times8$ $\mathbf{q}$-mesh~(29 $\mathbf{q}$ points in the 
irreducible wedge of the Brillouin Zone). Force constant interpolation 
is used to calculate phonon frequencies at arbitrary $\mathbf{q}$
vectors. The summation in Eq.~(\ref{eq:fph}) is evaluated on a 
$32\times32\times32$ $\mathbf{q}$-mesh.

The electronic free energy $F_{\rm ele}(V_i,T)$ is determined by using
the Mermin functional\cite{mermin} and Fermi-Dirac smearing. Similar to
getting $F_{\rm vib}(V_i,T)$, we first compute $F_{\rm ele}$ at every 
$50$~K from $50$~K to $5000$~K, then we fit them to a 4th order polynomial
\begin{equation}
F_{\rm ele}(V_i,T)=b_1(V_i)\cdot T+b_2(V_i)\cdot T^2+b_3(V_i)\cdot T^3
+b_4(V_i)\cdot T^4.
\label{eq:fele}
\end{equation}
Terms other than $b_2(V_i)T^2$ represent deviations from the lowest-order 
Sommerfeld expansion 
$F_{\rm ele}=-\frac{\pi^2}{6}N(E_{\rm F},V_i)(k_{B}T)^2$, 
where $N(E_{\rm F},V_i)$ is the electronic density of states at Fermi energy 
$E_{\rm F}$ and volume $V_i$. We find below $1000$ K, keeping only 
the quadratic term does not introduce much error. The influence of the higher 
order terms becomes prominent at high temperatures. At $2000$ K, the error 
reaches about $15$ \%. The fitted quadratic coefficient $b_2(V_i)$ differs 
from the Sommerfeld value $-\frac{\pi^2}{6}N(E_{\rm F},V_i)k_{B}^2$ by 
$5$~\%~($V_i$=$8.83$ \AA$^3$) to $15$~\%~($V_i$=$15.63$ \AA$^3$). It seems
the Sommerfeld expansion works better at high pressures, where the electronic
bands are more dispersive and $N(E_{\rm F})$ is smaller. We combine 
$F_{\rm ele}(V_i,T)$ with the static energy $U(V_i)$ from LAPW and the 
vibrational free energy $F_{\rm vib}(V_i,T)$ from the same pseudopotential to 
get the total free energy $F(V_i,T)$ at volume $V_i$. There are two popular 
parameterized forms to fit the total free energy $F(V,T)$, 4th order 
Birch-Murnahan\cite{birch}(BM) and Vinet.\cite{vinet} We find BM and Vinet 
are comparable in accuracy to fit the static and low temperature free energy, 
but BM yields much lower residual energies than Vinet for the high temperature 
results. Thus we use 4th order BM to get $F(V,T)$. Other thermodynamical 
properties are computed by finite difference.

All the pseudopotential calculations are carried out with the 
same plane-wave cutoff of $40$ Ry, charge-density cutoff of $480$ Ry,
and a shifted $16\times16\times16$ Monkhorst-Pack mesh. To determine the
convergence uncertainties of our results, we choose one 
volume~($V_i$ = $15.095$ \AA) and recompute its phonon frequencies at 
$T_{\rm ele}$ = 0 K, with a $24\times24\times24$ mesh and a higher plane-wave 
cutoff~($60$ Ry). The two sets of phonon frequencies differ by $0.5$ \% 
at most. The corresponding $F_{\rm vib}^{\rm QHA}$ differ by $0.07$ mRy at 
$2000$ K, $0.18$ mRy at $5000$ K. The influence of ETDV is much greater than 
the convergence uncertainties. For some modes phonon frequencies 
change by $10$ \% or more from $T_{\rm ele}$=$500$ K to $T_{\rm ele}$=$2000$ K.
The free energy correction $\Delta F_{\rm ETDV}$ is about $1$ mRy at $2000$ K.

%
%
\section{Summary of Previous Works}

Besides the two calculations\cite{xiang,proupin} mentioned in the 
introduction, which focus on the thermal EOS of platinum, there are some 
other papers related to this subject. 
Cohen {\it et al.}\cite{cohen-eos} computed the static EOS of platinum using
LAPW and PBE, and treated it as an example to discuss the accuracy of different 
EOS formations. They found Vinet fitted better than 3rd order BM. The accuracy
of 4th order BM and Vinet were comparable. Tsuchiya {\it et al.}\cite{taku-ep} 
computed the electronic thermal pressure~($P_{\rm ele}$) of Au and Pt using
LMTO and LDA. At $2200$~K, $P_{\rm ele}$ is $1.01$~GPa for Pt, while only 
$0.06$~GPa for Au. This is caused by the different $N(E_{\rm F})$ of the two 
metals. The small ETDV effect~($1$ \% change in phonon frequency from 
$T_{\rm ele}$= $0$ K to $3000$ K) observed on gold\cite{taku-au} is 
consistent with this picture. Wang {\it et al.}\cite{wang} used LAPW and 
an average potential method to determine the thermal contributions. Then 
they reduced the experimental shock Hugoniot and got the room temperature 
isotherm of Pt. This isotherm is very similar to that of Holmes {\it et al.}, 
in spite of the fact that in the latter case, thermal pressure is estimated 
semi-empirically. Ref.~\onlinecite{bercegeay} computed the static EOS of 
platinum using pseudopotentials with/without spin-orbit effects up to 
$150$ GPa. In the following section, we compare our calculations with these 
previous ones whenever appropriate.

On the experimental side, The reduced isothermal $P$-$V$-$T$ EOS from shock
wave experiments are widely used as primary pressure scales. At present they 
are also the only experimental sources for $P$-$V$-$T$ data at very high 
pressures. The shock Hugoniot of platinum was first obtained by using chemical 
explosives.\cite{mcqueen} The reduced room temperature isotherm was up to 
$270$~GPa. Holmes {\it et al.}\cite{holmes} went to higher compression 
ratio using a two-stage light-gas gun. The final shock Hugoniot is a 
combination of these two sets of data. In spite of the crucial role of the 
reduced shock EOS, its accuracy suffers from low precision in measurements, 
and theoretical simplifications made in the reducing process.\cite{chijioke,zha}
With the development of DAC and third-generation synchrotron light source, 
cross-checking different pressure scales became feasible. More accurate 
thermal EOS were obtained by using this method.\cite{dewaele,fei2004} 

Recently, Dorogokupets {\it et al.}\cite{dorogokupets} constructed a 
semi-empirical model to describe the thermal properties of Al, Au, Cu, Pt, 
Ta and W, The model contains about 20 parameters, which are fitted to the 
available experimental data on the heat capacity, enthalpy, volume, thermal 
expansivity, bulk modulus and shock Hugoniot. Based on this model they
reanalyzed the data in Ref.~\onlinecite{dewaele} up to 100 GPa. The resulting
EOS, which are consistent with the measured thermal properties, are believed
to be more accurate than the original in the corresponding pressure 
range.\cite{dewaele-comm} A simplified version of the model\cite{dor2} 
yields similar EOS at low pressures. However their high pressure 
extrapolations differ by 2.5~\% near 240 GPa. It will be interesting to 
use DFT to explore the EOS at very high pressures, which are still out of 
reach for the current DAC experiments. 

%
%
\section{Results and Discussions}
%
%
\subsection{Static Equation of State}
\label{sec:static}
\begin{figure}
\includegraphics[width=0.45\textwidth]{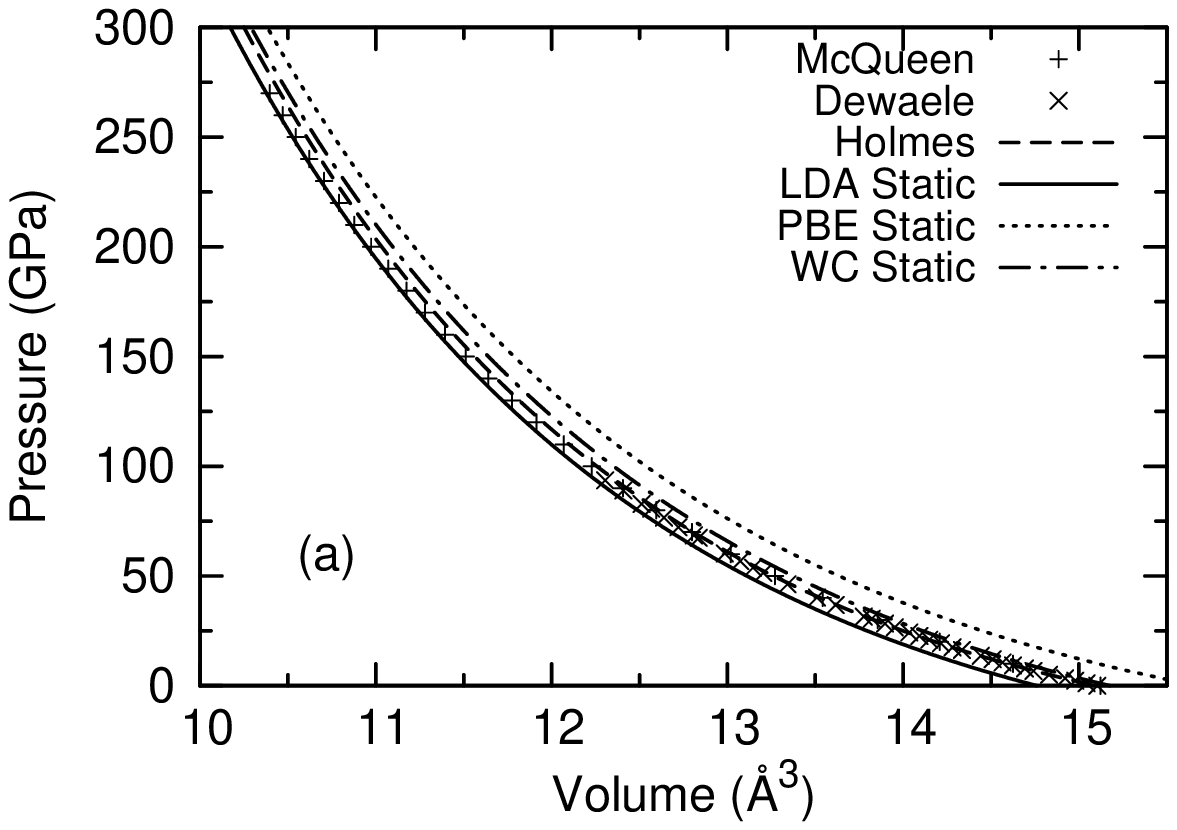}
\includegraphics[width=0.45\textwidth]{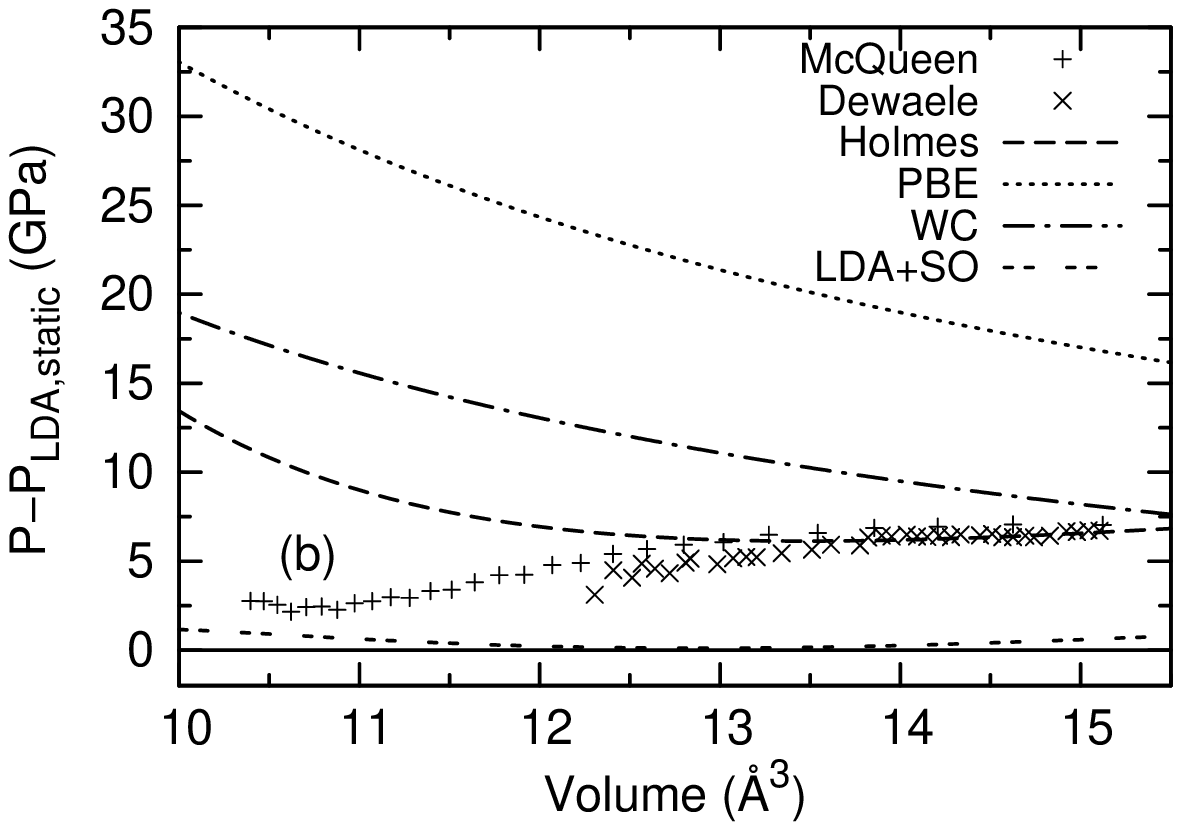}
\caption{Static EOS computed by the LAPW method using various exchange 
correlation functionals. (a) Pressure vs. volume curves. Spin-orbit effect 
is too small to be identified on this scale. (b) Pressure 
difference with respect to the static LDA EOS. In spite of the relative 
large change in EOS parameters, the actual pressure difference with/without 
spin-orbit effect is very small. Experimental data labeled as `McQueen' 
are from Ref.~\onlinecite{mcqueen}, `Dewaele' from Ref.~\onlinecite{dewaele}, 
`Holmes' from Ref.~\onlinecite{holmes}.}
\label{fig:p_v_static_exp}
\end{figure}

\begin{figure}
\includegraphics[width=0.72\textwidth]{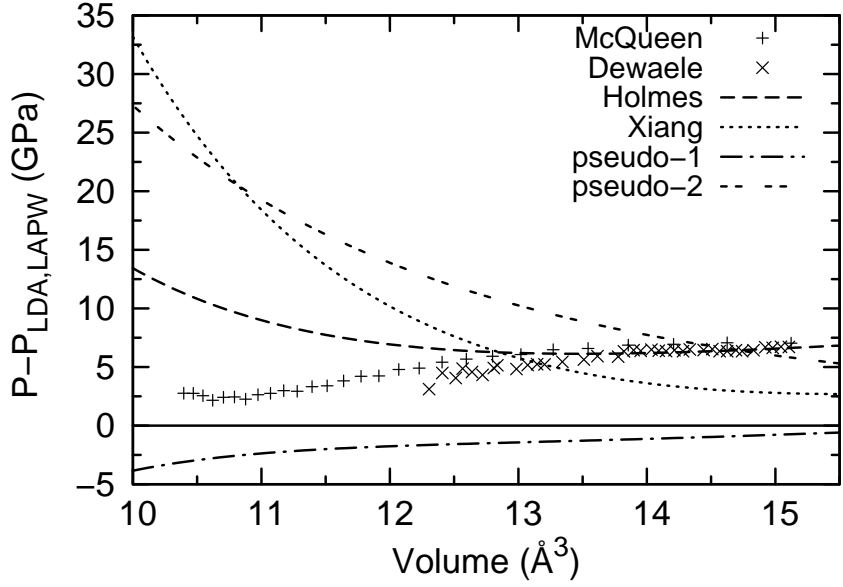}
\caption{Different LDA static EOS compared with the one computed by using 
LAPW. `Xiang' denotes the EOS from Ref.~\onlinecite{xiang}.}
\label{fig:p_v_static_lda}
\end{figure}

Before studying the EOS at finite temperature, we examine the static EOS
computed by using different exchange-correlation functionals, and compare
them with previous calculations. Excluding the thermal effects~(which amount 
to $\approx 2$~GPa at room temperature) helps to identify the origin of their 
differences. Fig.~\ref{fig:p_v_static_exp} shows the static pressure vs. 
volume relations using different exchange-correlation functionals.
The corresponding EOS parameters are listed in Table~\ref{tab:static}. 
The experimental data at room temperature are also included in the figure
to give a rough estimate of the difference. Comparing to the experiments, in 
the entire volume range LDA underestimates pressure while PBE overestimates. 
WC improves on PBE, but still overestimates. A detailed comparison between 
the calculated room temperature isotherms~(including the thermal effects) and 
the experimental data will be given in Sec.~\ref{sec:ambient}.  DFT has many 
different implementations, such as LAPW, LMTO, and various pseudopotentials. 
If the calculations are good, they should yield similar results. We compare 
our LDA calculations with previous ones in Fig.~\ref{fig:p_v_static_lda}. 
Two of our own pseudopotential calculations are included for comparison. 
One is the Vanderbilt pseudopotential that we use to compute the thermal 
effects, denoted as pseudo-1. The other is a Rappe-Rabe-Kaxiras-Joannopoulos 
pseudopotential from the PWSCF website~(Pt.pz-nd-rrkjus.UPF), denoted as 
pseudo-2. The static EOS predicted by pseudo-1 is similar to that of LAPW.
Their EOS parameters differ by no more than 0.5 \%. The previous 
overestimations of pressure are probably caused by the large cutoff 
radius and insufficient number of valence 
electrons~(Ref.~\onlinecite{proupin}), or another issue related to the 
negative electronic thermal pressure~(Ref.~\onlinecite{xiang}). 

%
%

%
%

%
%
\begin{table}
\caption{Static EOS parameters obtained from LAPW calculations and compared
with those in literature. Parameters from the pseudopotential 
calculations~(pseudo-1 and 2) are also listed. For convenience both Vinet 
and 4th order BM parameters are shown. $V_0$ denotes the equilibrium volume, 
$K_0$, $K'_0$ and $K''_0$ are the isothermal bulk modulus, the first and 
second derivative of the bulk modulus at $V_0$, respectively. Note their 
different fitting ranges: 
$0$-$550$ GPa~(this study), $0$-$1000$ GPa~(Ref.~\onlinecite{xiang}), 
$0$-$660$ GPa~(Ref.~\onlinecite{proupin}), 
$0$-$150$ GPa~(Ref.~\onlinecite{bercegeay}),
$0$-$350$ GPa~(Ref.~\onlinecite{cohen-eos}).
Ref.~\onlinecite{bercegeay} uses 3rd order BM EOS so the corresponding 
$K''_{0}$ are not listed.}
\begin{ruledtabular}
\begin{tabular}{cccccccc}
 & &Vinet & & & B-M && \\
 & $V_0$~(\AA$^3$) & $K_{0}$~(GPa) & $K'_{0}$ & $V_0$~(\AA$^3$) 
 & $K_{0}$~(GPa) & $K'_{0}$ 
 & $K''_{0}$~(GPa$^{-1}$)\\
\cline{1-1}\cline{2-4} \cline{5-8}
LDA&14.752&308.02&5.446&14.761&309.29&5.295&-0.02666\\
LDA+SO&14.784&301.17&5.533&14.785&301.13 &5.510 &-0.03214\\
LDA(pseudo-1)&14.719&308.69&5.423&14.726 & 309.61 & 5.295 &-0.02681\\
LDA(pseudo-2)&15.055&297.48&5.515&15.060 & 299.28 & 5.375 &-0.02873\\
LDA$^a$&14.90&300.9&5.814& & &\\
LDA$^b$&15.073&293&5.56& & &\\
LDA$^c$(HGH)& & & &14.82&305.99&5.32&--\\
LDA+SO$^{c,d}$(TM)& & & &15.2&291.18&5.35&-- \\
\hline
PBE&15.679&242.50&5.639&15.678&245.88&5.464&-0.03620\\
PBE+SO&15.751&231.97&5.762&15.754&229.96&5.850&-0.04932\\
PBE$^a$&15.77&243.3&5.866& & & & \\
PBE$^c$(HGH)& & & &15.59&250.85&5.65&--\\
PBE$^e$&15.69&248.9&5.43& & & & \\
\hline
WC&15.171&280.63&5.500&15.177&283.49&5.306&-0.02889\\
WC+SO&15.223&269.97&5.630&15.223&269.00&5.670&-0.03893\\
\end{tabular}
\end{ruledtabular}
\footnotetext[1]{Reference \onlinecite{xiang}.}
\footnotetext[2]{Reference \onlinecite{proupin}.}
\footnotetext[3]{Reference \onlinecite{bercegeay}.}
\footnotetext[4]{Reference \onlinecite{bercegeay-note}.}
\footnotetext[5]{Reference \onlinecite{cohen-eos}.}
\label{tab:static}
\end{table}

Platinum is a heavy element, and its electronic band structure is sensitive
to spin-orbit effect.\cite{corso} We find inclusion of the spin-orbit effect 
increases the equilibrium volume, no matter which exchange 
correlation functional is used. This tendency has also been observed by 
Bercegeay {\it et al.}\cite{bercegeay} in their pseudopotential calculations. 
However, the EOS parameters are not independent of each other. 
The variation of the equilibrium volume largely compensates that of the bulk 
modulus and the actual pressure difference is within 0.7~\% at high pressures.

Using pseudopotentials instead of the all electron LAPW may introduce 
error in computing phonon frequencies, especially at high pressures. Since 
lattice vibrations are closely related to the force/stress on the atoms, we 
estimate the error in phonon frequencies by analysing the error in static 
pressure. At high pressures~(150-550 GPa), the pressure difference between 
LAPW~(with LDA functional) and pseudo-1 is about $1.4$ \%. The error in 
phonon frequencies caused by using pseudo-1 is likely to be of the same order.
Since the influence of spin-orbit effect is half of the pseudopotential 
uncertainty, it is ignored completely in the following calculations.
%
%
\subsection{Phonon Dispersion and Its Electronic Temperature Dependence}
\label{sec:ph}
\begin{figure}
\includegraphics[width=0.45\textwidth]{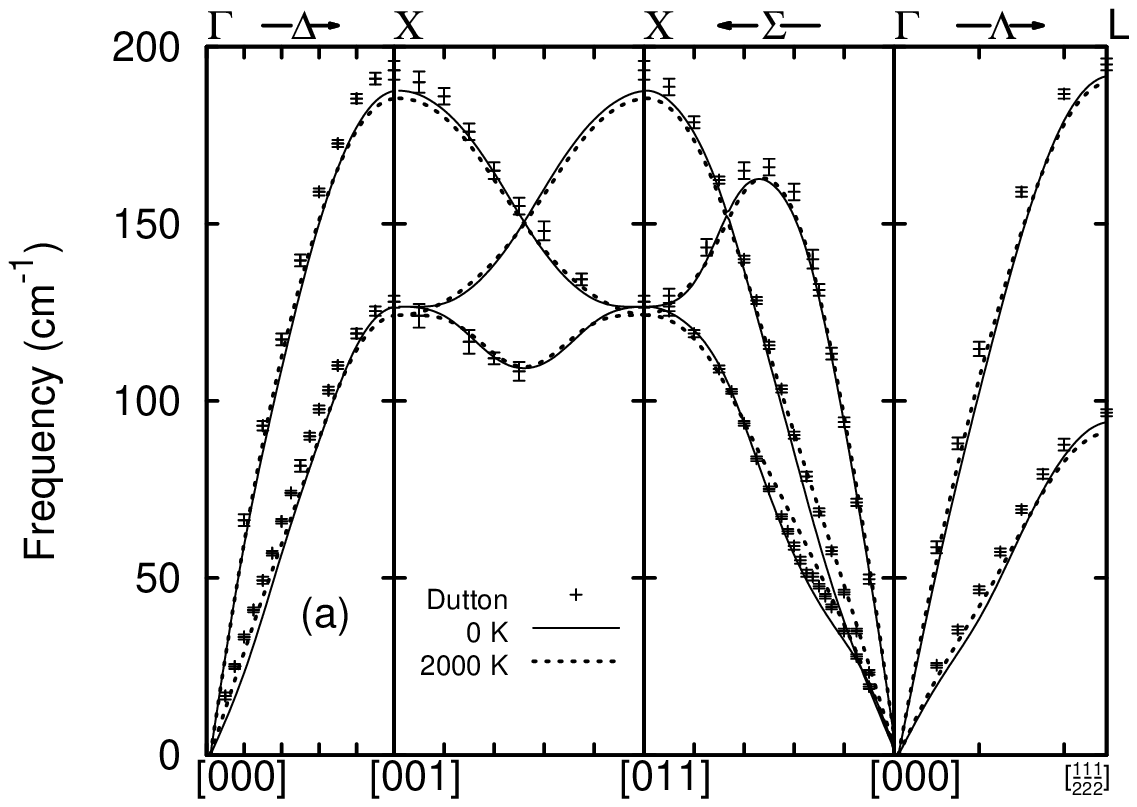}
\includegraphics[width=0.45\textwidth]{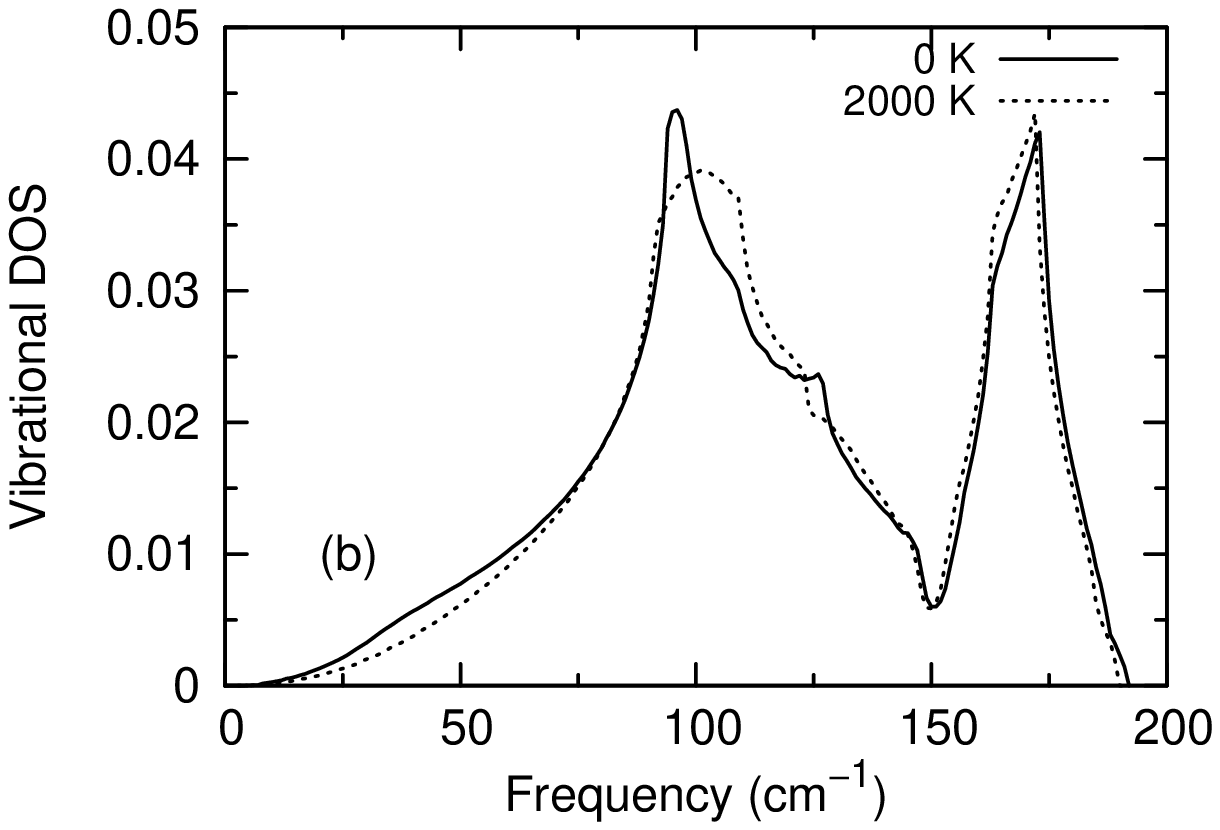}
\caption{(a) Phonon dispersion and (b) vibrational DOS at $a$=$7.4136$ a.u.. 
Solid line corresponds to $T_{\rm ele}$=0 K, dashed line corresponds to 
$T_{\rm ele}$=2000 K. They are calculated by using the LDA pseudopotential. 
Experimental data labeled as `Dutton' are from Ref.~\onlinecite{dutton}.}
\label{fig:phdisp}
\end{figure}

\begin{figure}
\includegraphics[width=0.45\textwidth]{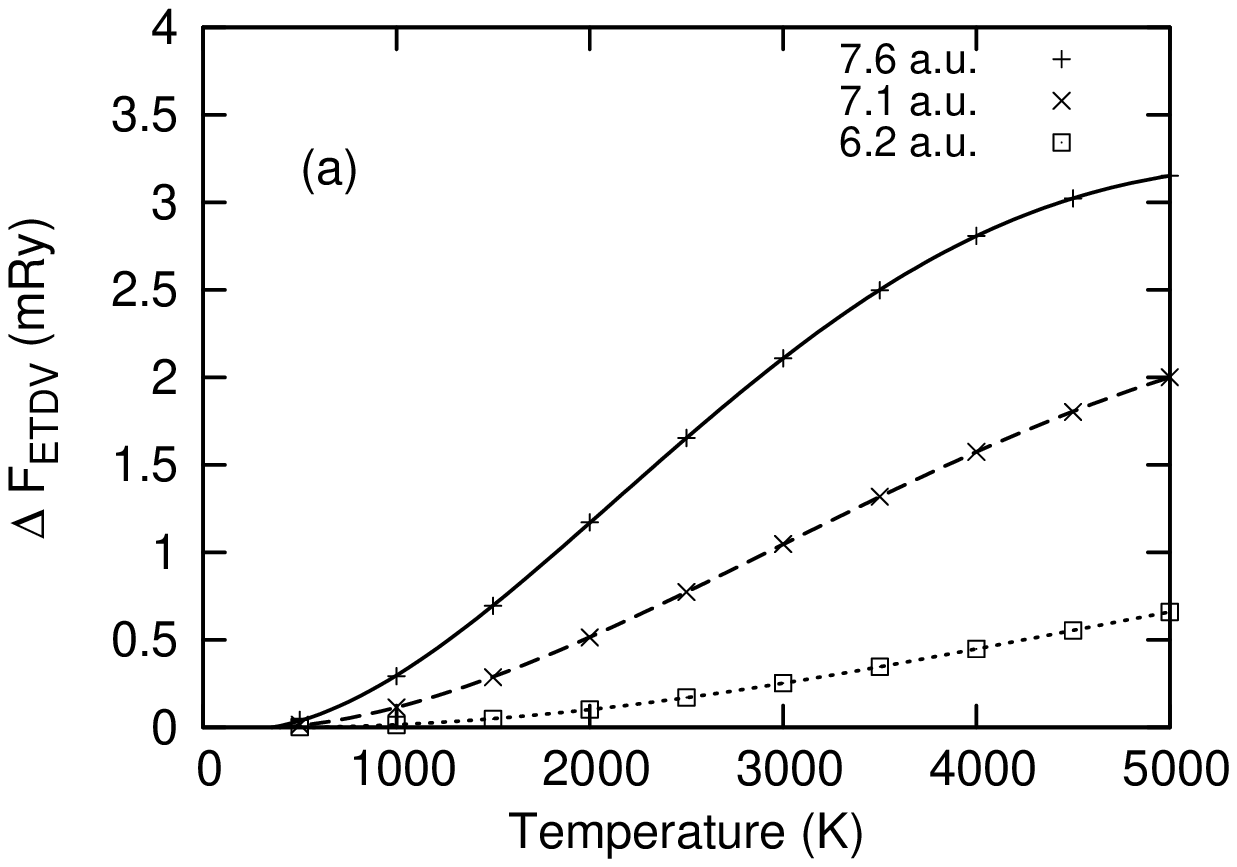}
\includegraphics[width=0.45\textwidth]{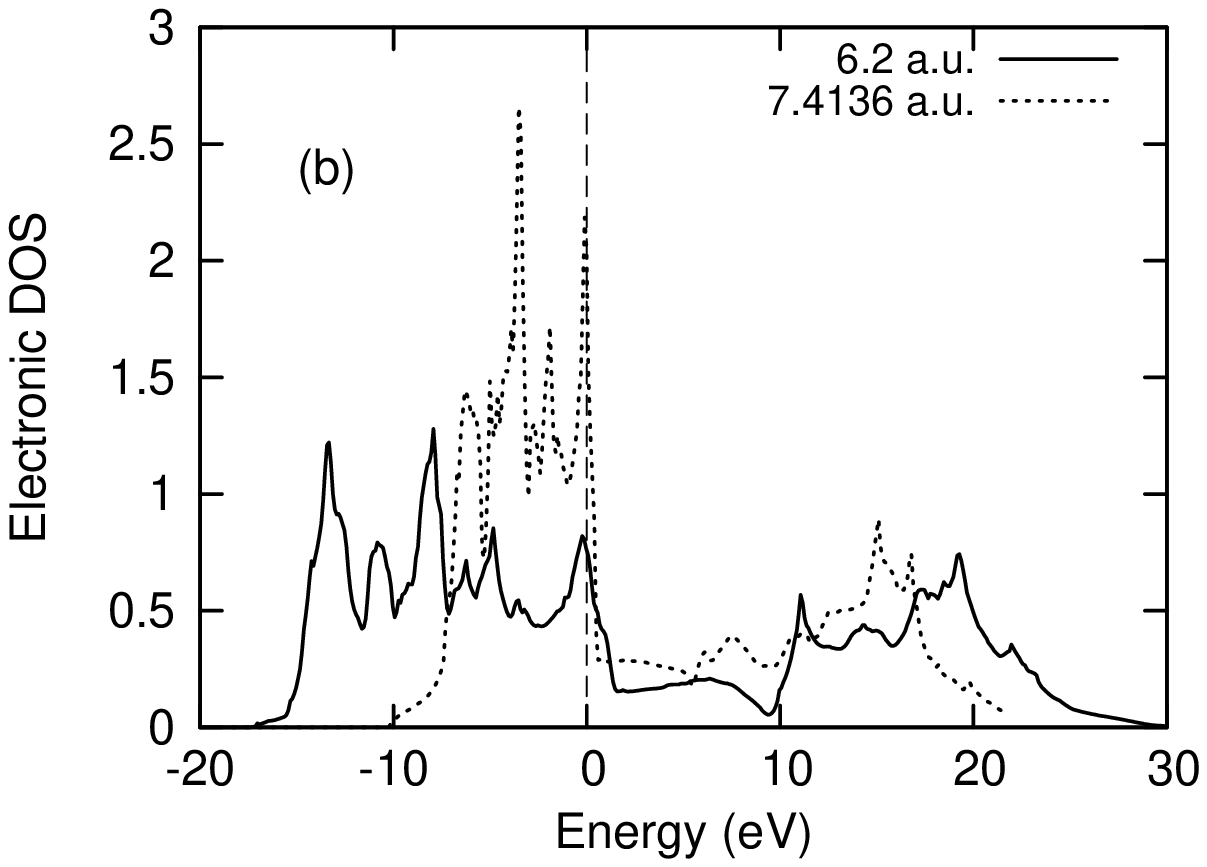}
\caption{(a) Corrections to the vibrational free-energy at various lattice
constants. (b) Volume dependence of the electronic density of states.}
\label{fig:phdf}
\end{figure}

Fig.~\ref{fig:phdisp} shows the phonon dispersions at the experimental
ambient condition lattice constant $a$=$7.4136$ a.u..\cite{dewaele} 
One is computed at $T_{\rm ele}$=$0$~K. The other at $T_{\rm ele}$=$2000$~K,
close to platinum's melting point at ambient pressure $2041.3$~K.\cite{kavner} 
The Kohn anomaly~(near $\mathbf{q}$=$\left[0,0.35,0.35\right]$) disappears 
when the electronic temperature is high, and the vibrational DOS varies 
noticeably. The corresponding corrections to the vibrational free energy,
$\Delta F_{\rm ETDV}(V_i,T)$, are shown in Fig.~\ref{fig:phdf}~(a). 
$\Delta F_{\rm ETDV}$ is always positive. As volume decreases, it diminishes 
and finally becomes negligible. ETDV originates in the thermal excitations
of the electrons near the Fermi surface, and the number of thermal excited 
electrons is proportional to $N(E_{\rm F})$ in the lowest-order Sommerfeld 
expansion. For smaller volumes, the electronic bands are more dispersive 
and $N(E_{\rm F})$ decreases, as shown in Fig.~\ref{fig:phdf}(b). ETDV 
diminishes accordingly.

%
%
%
%

\begin{figure}
\includegraphics[width=0.45\textwidth]{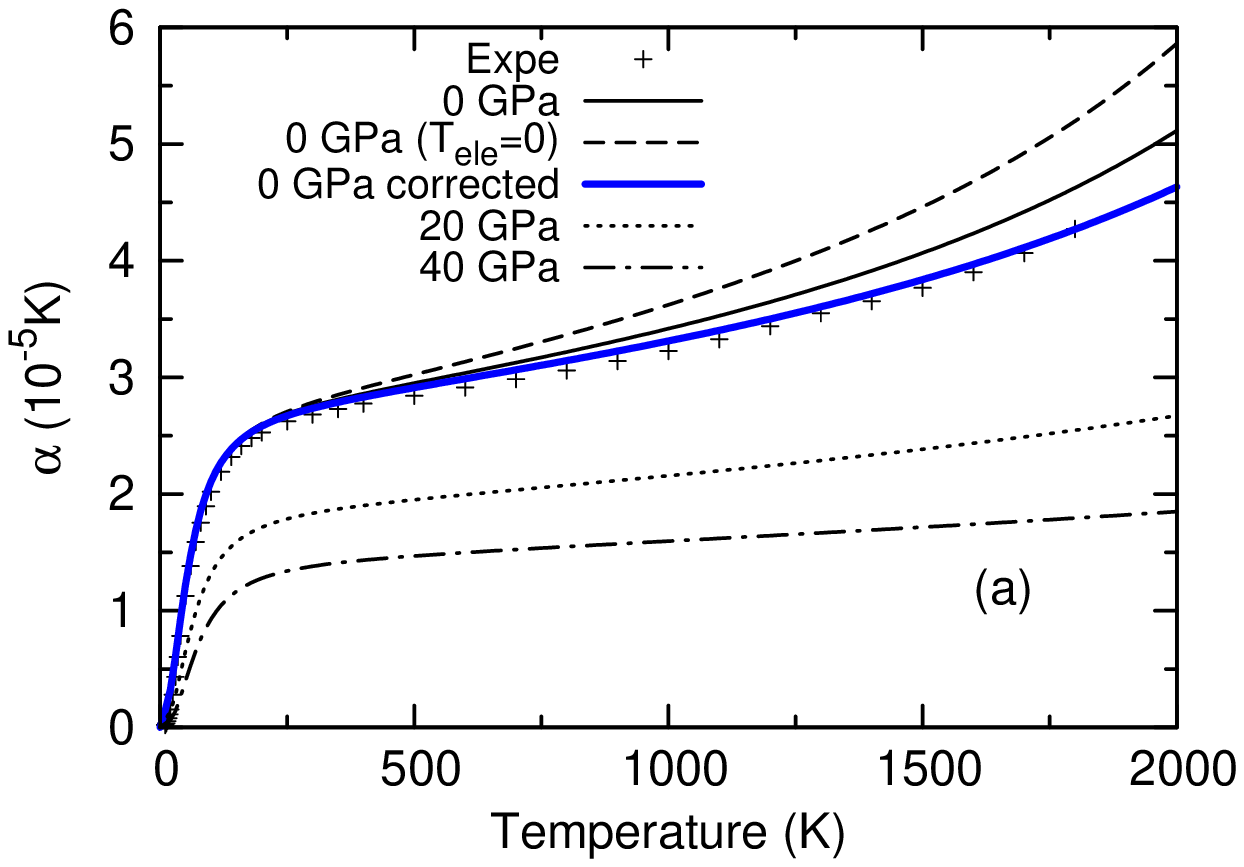}
\includegraphics[width=0.45\textwidth]{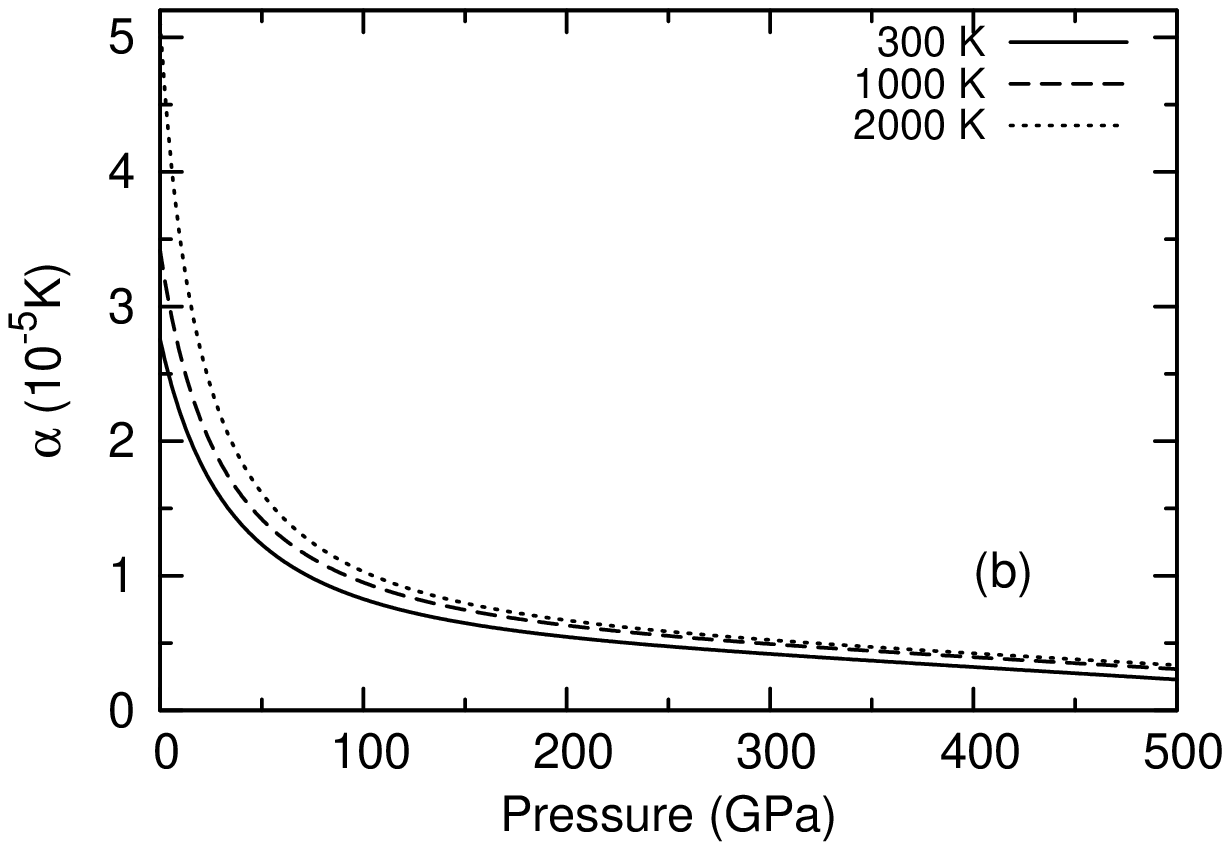}
\caption{Thermal expansivity as a function of (a) temperature and 
(b) pressure. Curves with label `(T$_{\rm ele}$=0)' represent properties 
computed without ETDV, i.e. computed from $F_{\rm vib}^{\rm QHA}$. Curves 
without this specification are the default ones computed with ETDV. `0 GPa
corrected' denotes the results obtained by adding a phenomenological 
correction to account for anharmonic phonon-phonon interactions and 
electron-phonon interactions. These corrected thermal data are used to 
construct the final thermal EOS in Table \ref{tab:isochor} and \ref{tab:para}.
The experimental data are from Ref.~\onlinecite{kirby}.}
\label{fig:alpha}
\end{figure}

\begin{figure}
\includegraphics[width=0.45\textwidth]{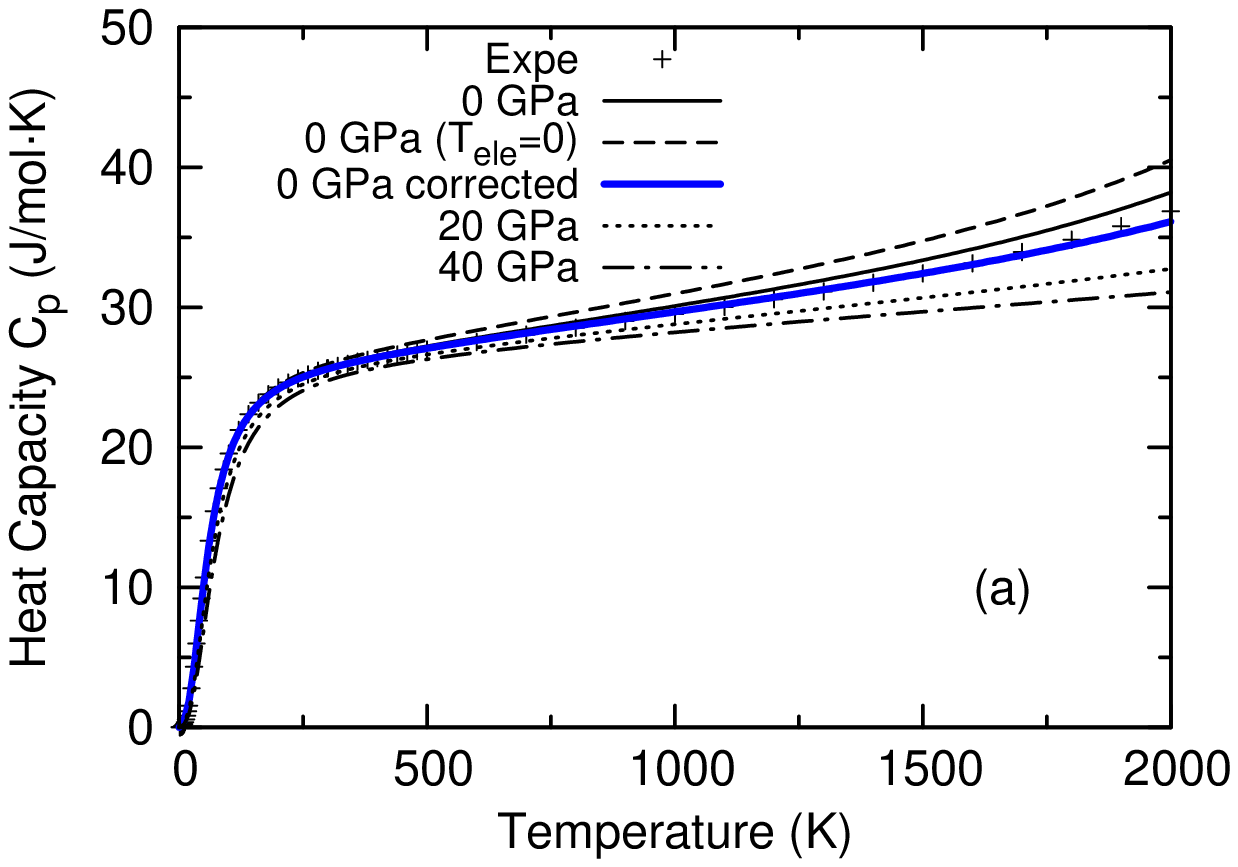}
\includegraphics[width=0.45\textwidth]{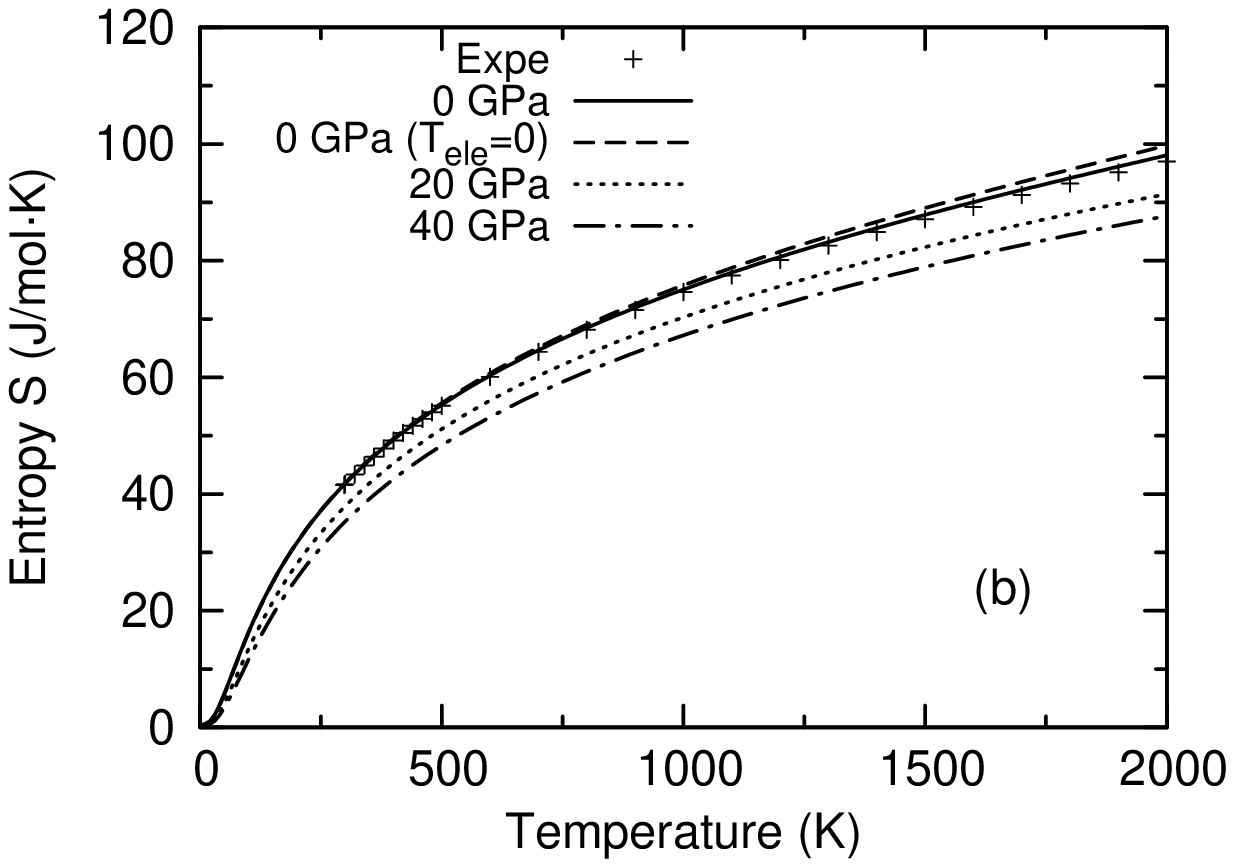}
\includegraphics[width=0.45\textwidth]{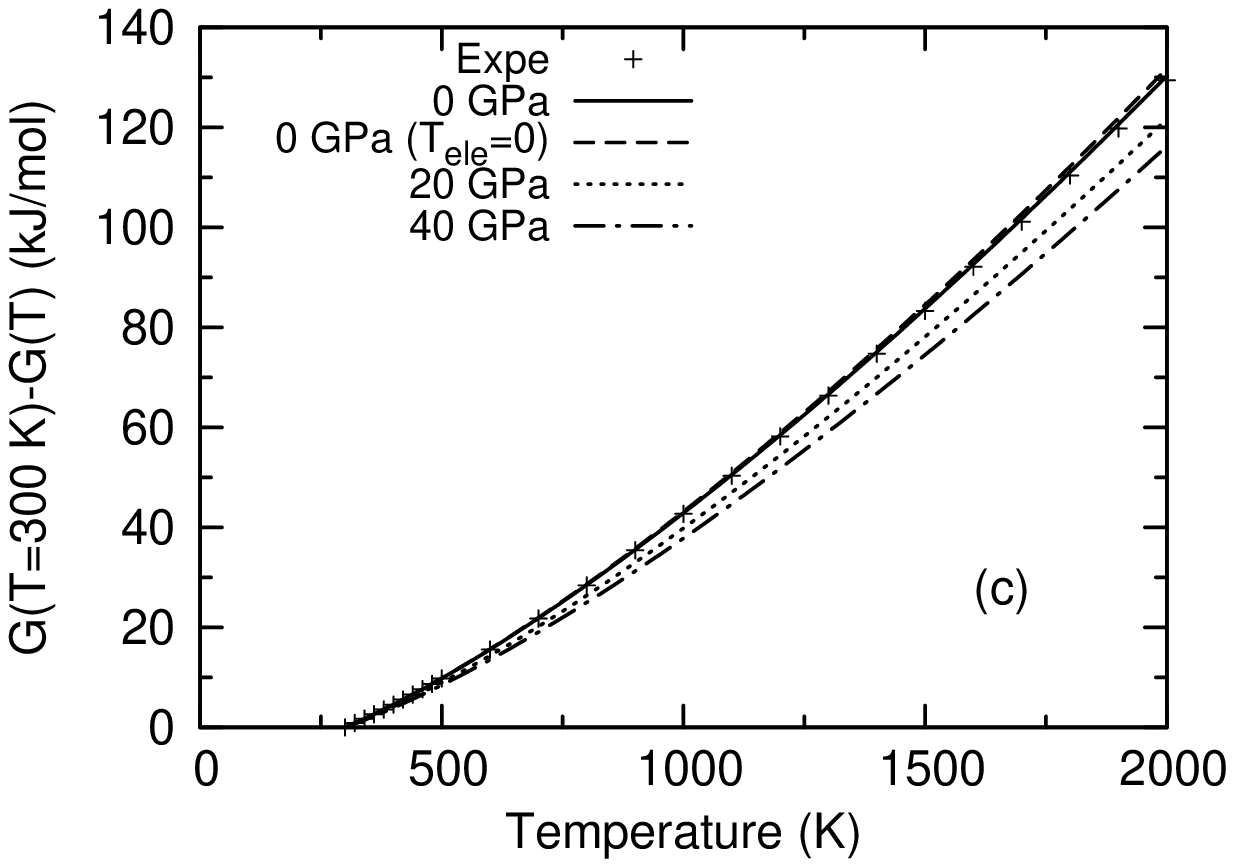}
\caption{Thermal properties of platinum. (a) Heat Capacity at constant 
pressure, (b) Entropy, (c) Temperature dependence of the Gibbs free energy at
constant pressure. The meanings of the labels are the same as above. On this
scale it is difficult to discern the improvements on entropy and Gibbs 
free energy, caused by including the phenomenological correction on anharmonic 
phonon-phonon interactions and electron-phonon interactions. Thus those data 
are not shown. The experimental data are from Ref.~\onlinecite{arblaster}.} 
\label{fig:cp-entropy-gibbs}
\end{figure}

Figure \ref{fig:alpha} and \ref{fig:cp-entropy-gibbs} show the volume 
thermal expansion coefficient $\alpha$, heat capacity at constant 
pressure $C_{P}$, entropy $S$, and the temperature-dependent part of the
Gibbs free energy $\Delta G(P,T)$=$G(P,T)$-$G(P,T=300 K)$. Including ETDV 
removes about half of the discrepancies between experiments and calculations
based on normal QHA. The remaining small differences between 
theory and experiment are attributed to anharmonic phonon-phonon 
interactions\cite{cowley} and electron-phonon interactions.\cite{allen} 
These two effects are of the same order of magnitude\cite{allen} as 
$F_{\rm ele}$ for metals, but explicit perturbative calculations to determine
their magnitudes are computational demanding and beyond the scope of the 
current paper. We notice DFT calculations based on QHA describe well the 
thermal properties of other metals, such as gold,\cite{taku-au} 
silver,\cite{xie} copper,\cite{narasimhan} up to melting point. This is in 
contrast with ionic crystals like MgO, where there are large deviations from 
QHA at high temperatures. It is possible that the effects of anharmonic 
phonon-phonon interaction and electron-phonon interaction tend to
cancel each other in these metals. Further work is needed to clarify this
issue.

%
%
%
%

%
%
\subsection{Room Temperature Isotherms}
\label{sec:ambient}

\begin{figure}
\includegraphics[width=0.45\textwidth]{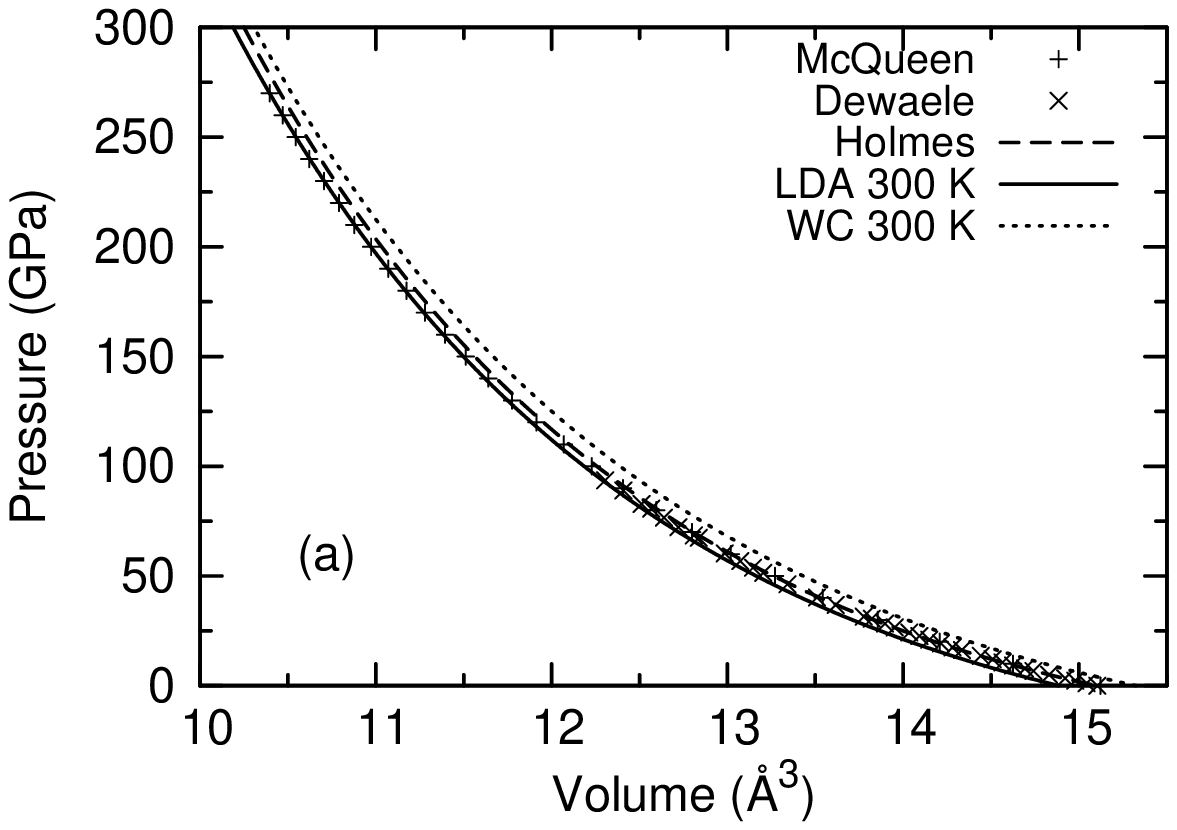}
\includegraphics[width=0.45\textwidth]{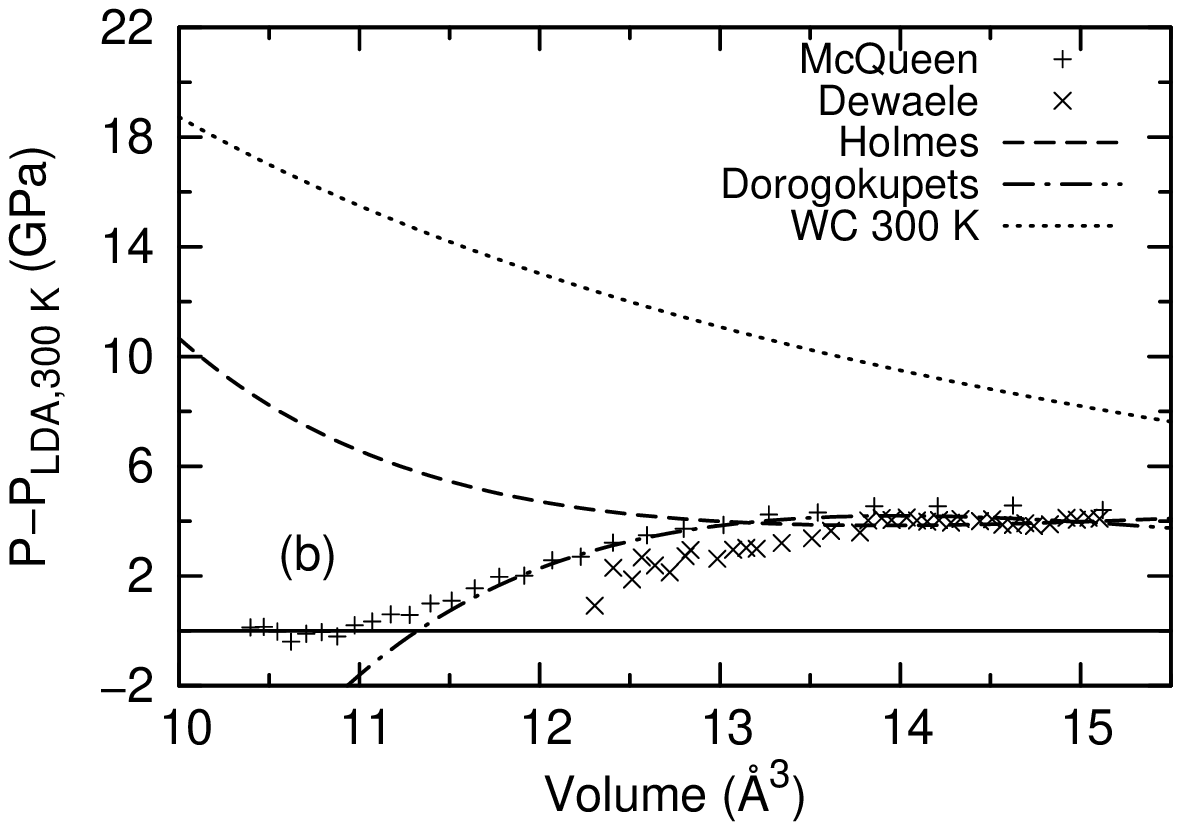}
\caption{ 300 K isotherms. (a) pressure vs. volume. 
(b) pressure difference. `Dorogokupets' denotes EOS from 
Ref.~\onlinecite{dorogokupets}, which is very close to the reduced shock 
EOS~(McQueen) near $100$ GPa. PBE is not plotted as its EOS is way off.}
\label{fig:isotherm}
\end{figure}

By fitting the total Helmholtz free energy at $300$~K, we get the theoretical 
$300$~K isotherms, as shown in Fig.~\ref{fig:isotherm}. Their parameters are 
listed in Table~\ref{tab:isotherm}. In the low pressure range, the LDA 
isotherm and the experimental data are almost parallel. As pressure 
increases, they start to merge. It seems LDA works better at high
pressures. Regarding to EOS parameters, LDA gives equilibrium 
volumes closest to the experiments, WC yields closest bulk 
modulus~($K_0$) and the derivative of the bulk modulus~($K'_0$). 
Some people\cite{dewaele,bercegeay} prefer to compare 
pressures from two EOS~(labeled as EOS-I and EOS-II) at the same compression, 
i.e. the same value of $V/V_0$. $V_0$ is the corresponding equilibrium volume, 
$V_{0,{\rm I}}$ for EOS-I, $V_{0,\rm II}$ for EOS-II. Such comparisons can 
give favorable agreement when $K_0$ and $K'_0$ of EOS-I are close to those
of EOS-II, even $V_{0,\rm I}$ and $V_{0,\rm II}$ are quite 
different.\cite{dewaele} As mentioned before, the EOS parameters 
are not independent of each other. It can be fortuitous that $K_0$ and 
$K'_0$ agree well. Judged from pressure vs. volume relation, LDA is the 
optimal functional for platinum. It is worth noting the LDA~(HGH) 
pressure vs. volume relation reported in Ref.~\onlinecite{bercegeay} is 
similar to those obtained in this study. However, 
Ref.~\onlinecite{bercegeay} presents data in volume vs. 
compression, and concludes LDA overestimates pressure by $8$ GPa near 
$100$ GPa. In fact, although $K_{0,\rm LDA}$~($291$ GPa from this 
study) is much larger than $K_{0,\rm expe}$~($273.6$ GPa from 
Ref.~\onlinecite{dewaele}), the bulk modulus computed at the experimental 
equilibrium volume $V_{0,\rm expe}$~(15.095 \AA$^3$) is $270$~GPa, quite 
close to $K_{0,\rm expe}$. Thus when plotted in pressure vs. volume, the 
isotherm computed by LDA is nearly parallel with the experimental data in 
the low pressure range.

%
%
%

%
%
\begin{table}
\caption{EOS parameters of the theoretical $300$~K isotherms, compared with 
the experiments. $V_{0,\rm expe}$ is $15.095$~\AA$^3$. 
Pressure range: $0$-$550$ GPa~(this study), 
$0$-$660$ GPa~(Ref.~\onlinecite{holmes} and \onlinecite{proupin}), 
$0$-$94$ GPa~(Ref.~\onlinecite{dewaele} and \onlinecite{dorogokupets}), 
$0$-$270$ GPa~(Ref.~\onlinecite{mcqueen}).}
\begin{ruledtabular}
\begin{tabular}{cccccccc}
 & &Vinet & & & B-M && \\
 & $V_0$~(\AA$^3$) & $K_{0}$~(GPa) & $K'_{0}$ & $V_0$~(\AA$^3$) 
 & $K_{0}$~(GPa) & $K'_{0}$ 
 & $K''_{0}$~(GPa$^{-1}$)\\
\cline{1-1}\cline{2-4} \cline{5-8}
LDA&14.884&291.25&5.547&14.886&291.65&5.496&-0.03232\\
LDA~($V_{0,\rm expe}$)&&269.96&5.640&&269.91&5.626&-0.03730\\
LDA$^a$&15.188&281&5.61& & &\\
PBE&15.864&225.55&5.751&15.866&225.34&5.741&-0.04709\\
WC&15.322&263.93&5.601&15.325&264.72&5.530&-0.03580\\
\hline
Holmes$^b$&15.10&266&5.81 & & & &\\
Dewaele$^{c}$&15.095&273.6&5.23& & & &\\
Dorogokupets$^{d}$&15.095&276.07&5.30& & & &\\
McQueen$^{e}$& & & &15.123&277.715&4.821&-0.01379\\
\end{tabular}
\end{ruledtabular}
\footnotetext[1]{Reference \onlinecite{proupin}.}
\footnotetext[2]{Reference \onlinecite{holmes}.}
\footnotetext[3]{Reference \onlinecite{dewaele}. When $K_0$ is set to 
$277$ GPa, the value measured by ultrasonic experiments, 
$K'_0$ equals $5.08$ GPa.}
\footnotetext[4]{Reference \onlinecite{dorogokupets}, improved analysis using 
data from Ref.~\onlinecite{dewaele}.}
\footnotetext[5]{Reference \onlinecite{mcqueen}, Fitted from the tabulated shock reduced isotherm at $293$~K.}
\label{tab:isotherm}
\end{table}
%
%
\subsection{Thermal EOS of Platinum for Pressure Calibration}
In the previous sections the thermal properties of platinum is discussed 
from a pure theoretical point of view. We have computed the static lattice 
energy $U(V)$ using LAPW, and found spin-orbit interactions are not important 
in determining the EOS of platinum. We have used QHA to calculate the 
vibrational free energy $F_{\rm vib}(V,T)$, and found including ETDV
improves the agreement on the thermal properties. We have calculated the
electronic free energy $F_{\rm ele}(V,T)$ using Mermin functional. The
resulting thermal properties, e.g. the temperature-dependent part of the
Gibbs energy $\Delta G(P,T)$, are close to the experimental data at $0$ GPa, 
The room temperature isotherm computed by LDA merges to the reduced shock
data at high pressures, indicating LDA works better at high pressures.

Based on these DFT results and all the available experimental data, we try
to construct a consistent $P$-$V$-$T$ EOS of platinum up to $550$ GPa 
and $5000$ K. To reach this goal, first we need to include the physical 
effects which are missing in our original model. A phenomenological 
term\cite{dor2} $\Delta F_{\rm corr}(V,T)=-\frac{3}{2}k_{\rm B}a(V/V_0)^mT^2$ 
is added to the total Helmholtz energy to account for the anharmonic 
phonon-phonon interactions and electron-phonon interactions, where $V_0$ is 
the volume of a primitive cell at ambient condition~($V_0=15.095$ \AA). The 
quadratic temperature dependence comes from the lowest order perturbation at 
high temperatures. $a$ and $m$ are two parameters to be fitted. We find 
setting $a$ equals $10^{-5}$ K$^{-1}$, $m$ equals $7$ yields good
agreement between theory and experiments on $\alpha$, $C_P$ and 
$\Delta G(P,T)$ at $0$ GPa, as illustrated in Figs.~\ref{fig:alpha}(a) and
\ref{fig:cp-entropy-gibbs}(a). The contribution to thermal pressure can be 
estimated by differentiating $\Delta F_{\rm corr}(V,T)$ with respect to 
volume. At $2000$ K, $\Delta P_{\rm th}$ is $0.38$ GPa when $V$ equals $V_0$,
$0.2$ GPa when $V$ equals $0.9V_0$. 

Having obtained accurate $\Delta G(P,T)$, the next step is to get reliable
$G(P,300~{\rm K})$. We choose the room temperature EOS developed by 
Dorogokupets {\it et al.}\cite{dorogokupets} as our reference below $100$ GPa.
It has been cross checked with other pressure scales, and is likely to be
more accurate than the reduced shock data of Ref.~\onlinecite{mcqueen} in this
pressure range. On the other hand, extrapolating an EOS fitted at low 
pressures to higher range can be dangerous. We assume LDA works better at
high pressures, and the difference between the exact~(obtained in an ideal,
very accurate experiment) and LDA isotherm approaches to zero as pressure 
increases. 

\begin{figure}
\includegraphics[width=0.72\textwidth]{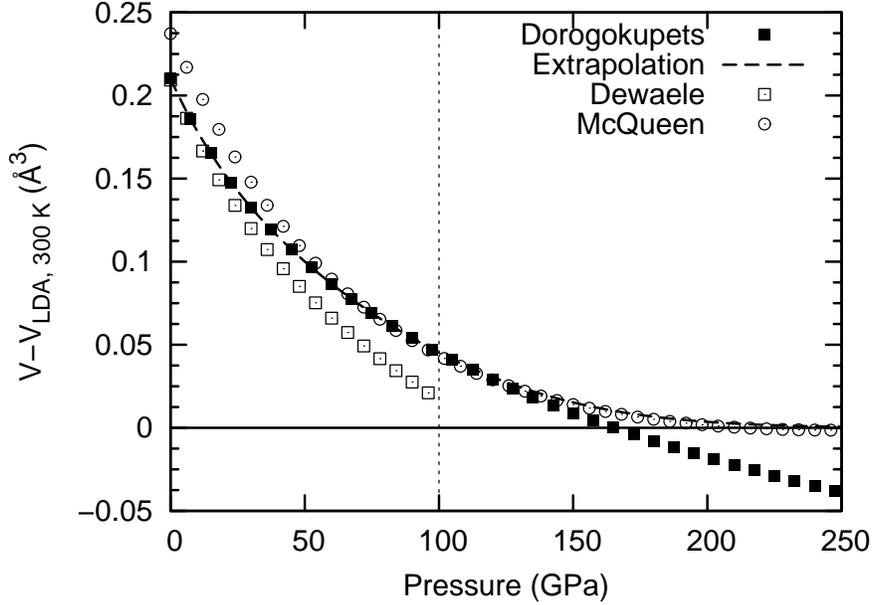}
\caption{Volume Correction to the theoretical isotherm at $300$~K.
We use 
$\Delta V(P)= 0.1215\cdot\exp(-P/34.0846)+0.0885\cdot\exp(-(P/109.989)^2)$ 
to fit the volume difference in the range $P < 100$ GPa. The exponential
functional form guarantees that it approaches zero at high pressures. It
happens that this extrapolation agrees well with McQueen's reduced shock
data.}
\label{fig:vol-corr}
\end{figure}

We compare Dorogokupets's EOS\cite{dorogokupets}~($V_0$=15.095 \AA$^3$, 
$K_0$=276.07 GPa, $K'_0$=$5.30$ in Vinet form) with our room temperature 
isotherm computed by LDA~($V_0$=14.884 \AA$^3$, $K_0$=291.25 GPa, 
$K'_0$=$5.547$ in Vinet form). The volume difference between these two 
at each pressure $\Delta V(P) = V_{\rm expe,300~{\rm K}}(P)-
V_{\rm LDA,300~{\rm K}}(P)$ is shown in Fig.~\ref{fig:vol-corr}. 
Since $\Delta V(P)$ decreases rapidly as pressure increases, we use 
exponentially decaying functions to fit and extrapolate. We correct the
calculated room temperature Gibbs energy $G_{\rm LDA}(P,300~{\rm K})$ by 
setting $G_{\rm corr}(P,300~{\rm K})=G_{\rm LDA}(P,300~{\rm K})
+\int_{0}^{P}\!\Delta V(P)dP$. The isotherm derived from 
$G_{\rm corr}(P,300~{\rm K})$ coincides with Dorogokupets' EOS below $100$ GPa. 
the upper limit of their fitting. Above $250$~GPa, $\Delta V(P)$ is 
almost zero, and the isotherm derived from $G_{\rm corr}(P,300~{\rm K})$ 
is the same as the uncorrected one. The uncertainty due to volume 
extrapolation in the intermediate region~($100$ GPa to $250$ GPa) is estimated 
from bulk modulus to be less than $1$~GPa. It is worth noting that the 
established EOS of platinum are quite quite similar to each other below 
$100$ GPa, as shown in Fig.~\ref{fig:isotherm}(b). Choosing a different 
reference such as the one in Ref.~\onlinecite{dewaele} will only change the 
results near $100$ GPa by $1$ GPa.

%
%

\begin{figure}
\includegraphics[width=0.45\textwidth]{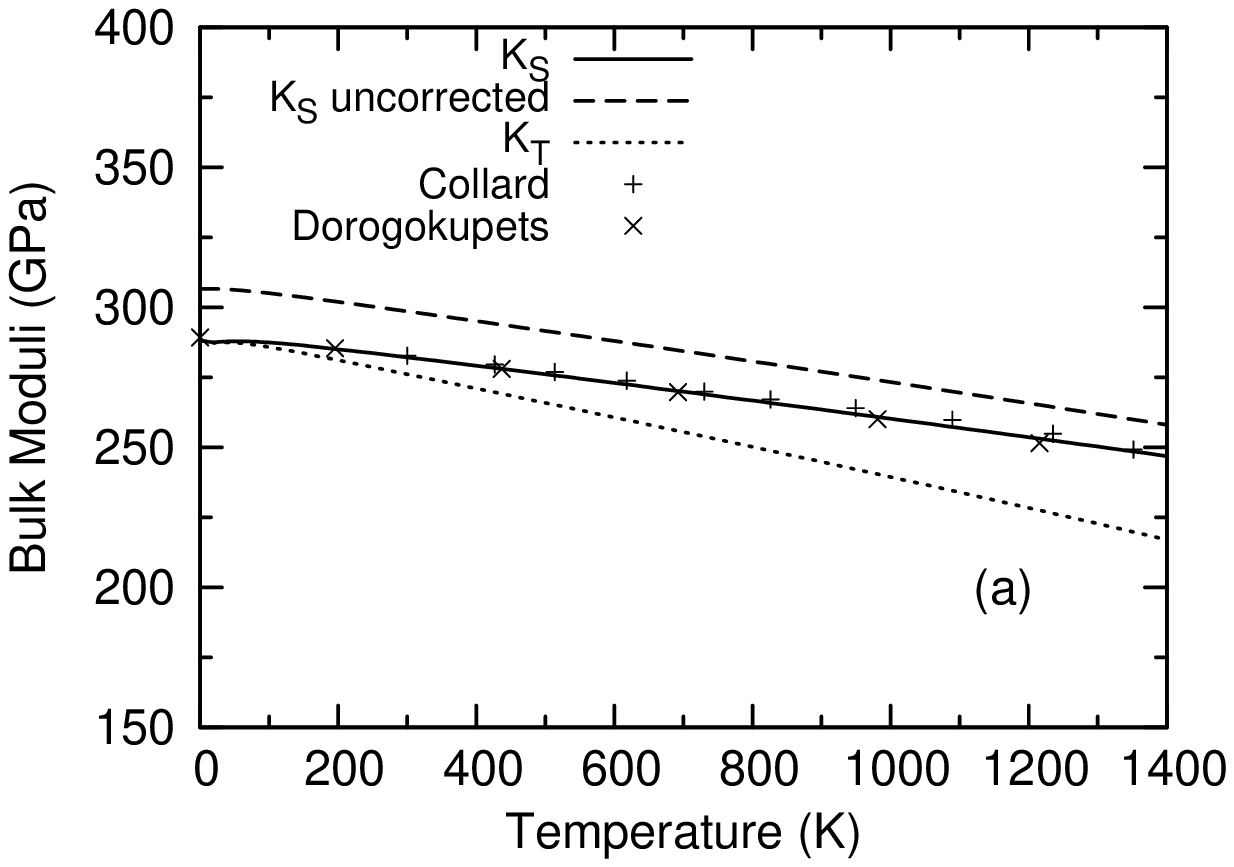}
\includegraphics[width=0.45\textwidth]{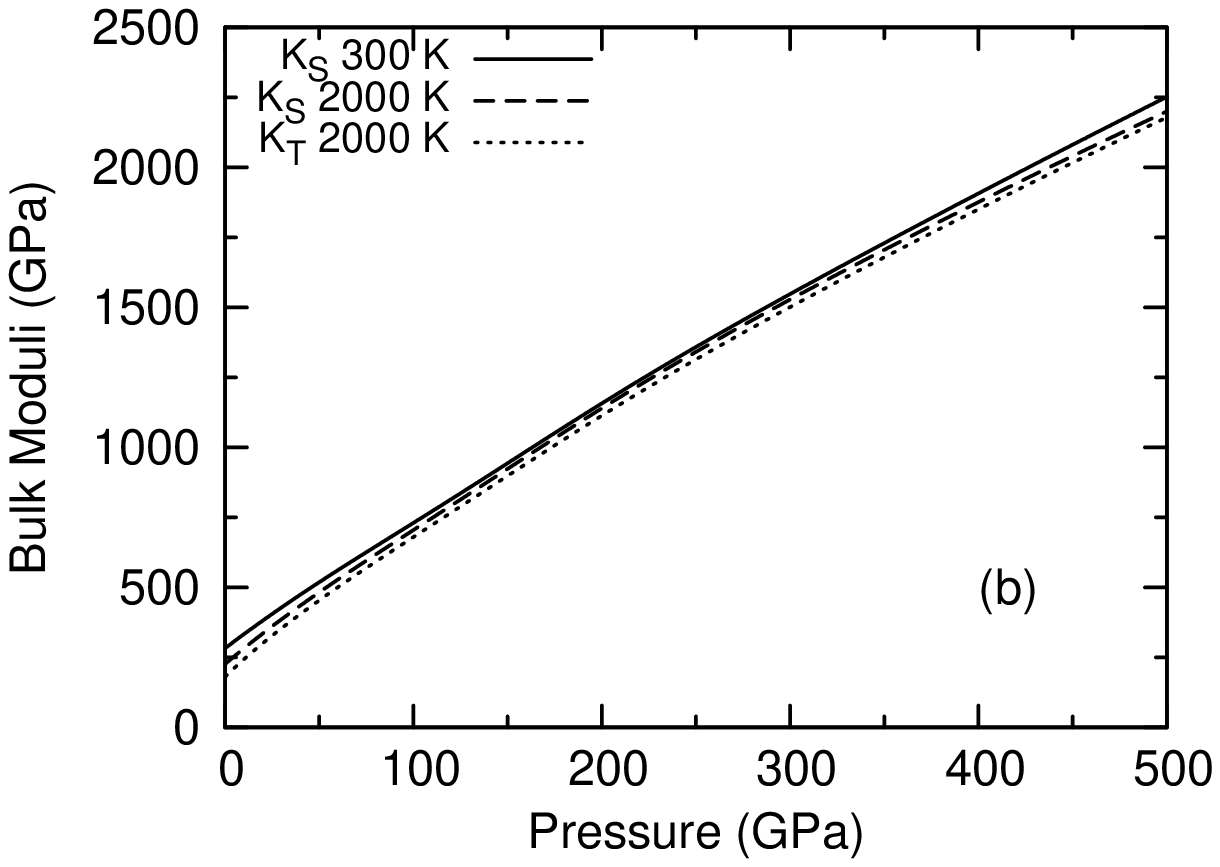}
\caption{ (a) Bulk moduli at $0$ GPa. `Uncorrected' denotes the raw DFT 
results without any correction. As noted in Ref.~\onlinecite{dorogokupets},
the experimental data in Ref.~\onlinecite{collard} is inconsistent. The data 
points we show here are digitized from the graphs in 
Ref.~\onlinecite{collard}~(denoted as `Collard') 
and \onlinecite{dorogokupets}~(denoted as `Dorogokupets') respectively. 
$K_S$ deduced from the corrected Gibbs free energy agrees well with the one 
computed from empirical models in Ref.~\onlinecite{dorogokupets}. (b) Bulk 
moduli as a function of pressure.}
\label{fig:ks}
\end{figure}

We combine $G_{\rm corr}(P, 300~{\rm K})$ with the temperature-dependent 
part of the Gibbs energy $\Delta G(P,T)$, and get the corrected Gibbs energy 
$G_{\rm corr}(P,T)$ at temperature $T$. From $G_{\rm corr}(P,T)$ we derive all 
the other thermodynamical properties. Thermal properties like $\alpha$, 
$C_P$, $S$, which depend on the temperature derivatives of the Gibbs 
energy, are not affected by changing $G(P, 300~{\rm K})$. In contrast, the 
isothermal bulk modulus $K_T$ and adiabatic bulk modulus $K_S$ will be 
influenced, as shown in Fig.~\ref{fig:ks}.

%
%

\begin{figure}
\includegraphics[width=0.45\textwidth]{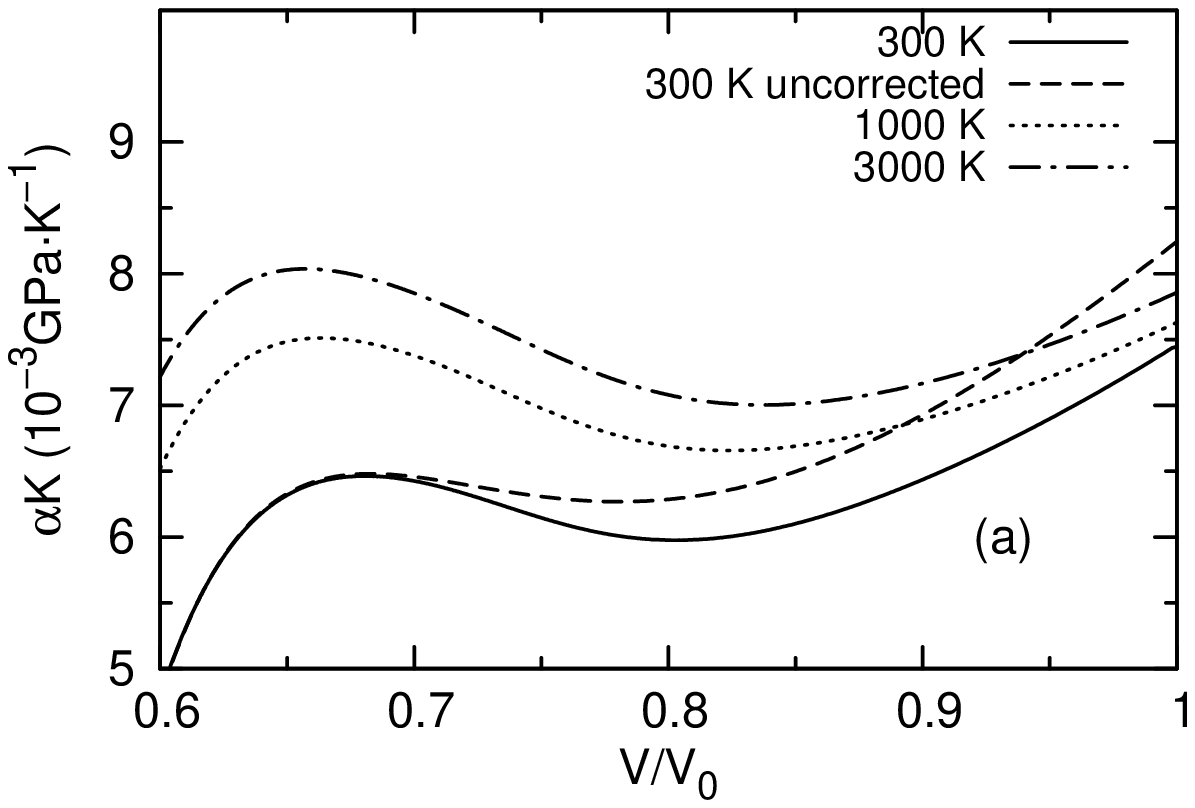}
\includegraphics[width=0.45\textwidth]{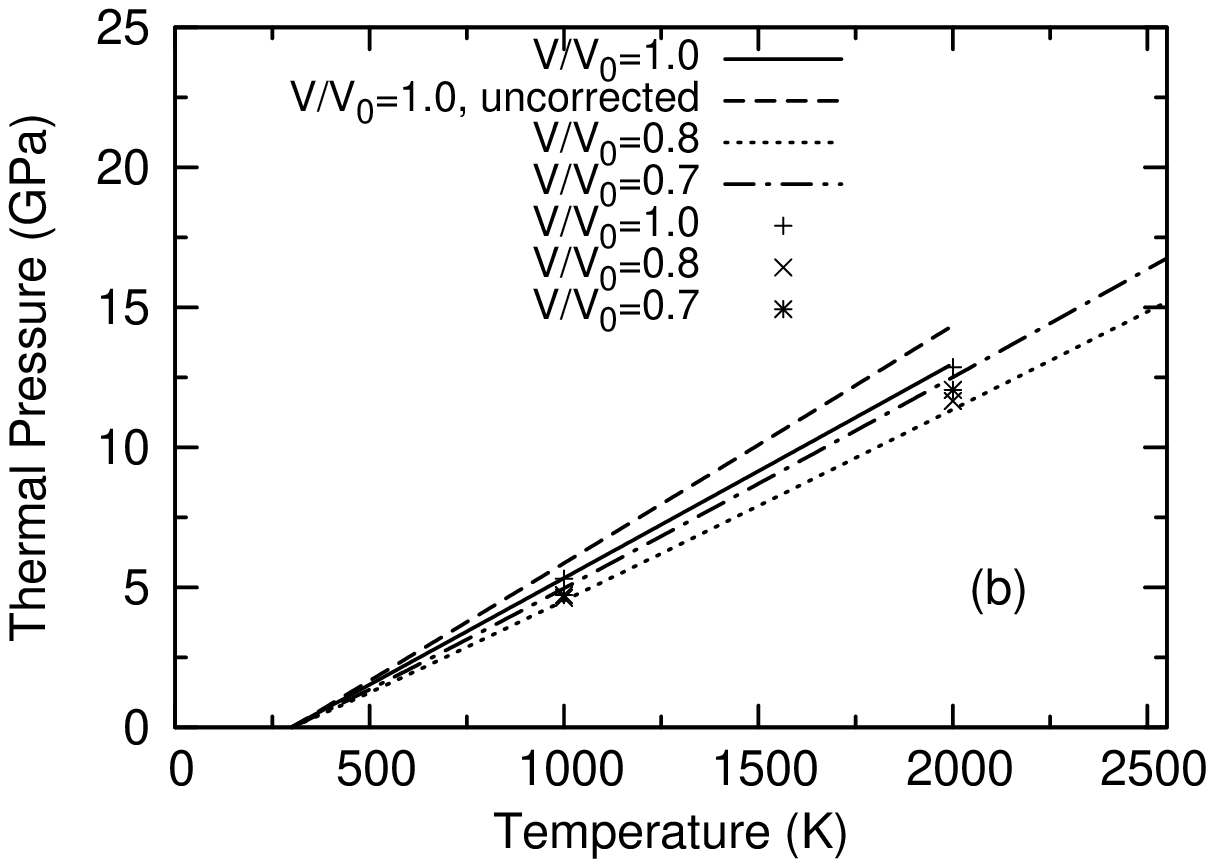}
\caption{ (a) temperature derivative of $P_{\rm th}$, 
$\alpha K_{T}$=$\frac{\partial P_{\rm th}}{\partial T}$ before/after 
corrections. (b) Thermal pressure $P_{\rm th}$ at different 
$V/V_0$, where $V_0$ is the experimental volume at ambient 
condition~(15.095 \AA$^3$). Points are the thermal pressures from 
Ref.~\onlinecite{dorogokupets}.}
\label{fig:thp}
\end{figure}

After corrections, both thermal expansivity and bulk modulus agree with
the experiments well. We expect the product $\alpha K_T$ 
to be accurate. Integrating  $\alpha K_T$ we get the thermal pressure,
$P_{\rm th}(V,T)=P(V,T)-P(V,T_0)=\int_{T_0}^T\alpha K_{T}dT$.  The
calculated $\alpha K_T$ and $P_{\rm th}$, before and after corrections, are
shown in Fig.~\ref{fig:thp}. $P_{\rm th}(V,T)$ is often assumed to be 
independent of volume and linear in temperature, i.e. $\alpha K_T$ is a 
constant.  Ref.~\onlinecite{mcqueen} assumes the thermal 
energy $E(T)=3k_{B}T$, the thermal Gr\"uneisen parameter 
$\gamma=\gamma_{0}V/V_0$, where $\gamma_{0}$=$2.4$, and $V_0$=$15.123$ \AA$^3$.
The thermal pressure is obtained from Mie-Gr\"uneisen relation
\begin{equation}
P_{\rm th}(V,T)=\frac{E(T)\gamma(V)}{V}=\frac{3k_{B}\gamma_{0}}{V_0}\cdot T
=6.57\times10^{-3}\cdot T~({\rm GPa}).
\label{eq:mie-gruneisen}
\end{equation}
In Ref.~\onlinecite{holmes}, $\alpha K_T$ is estimated to be 
$6.94 \times 10^{-3}$~GPa/K. Both values lie within the variation of the 
calculated $\alpha K_{T}$, as shown in Fig.~\ref{fig:thp}(a). We find that 
$\alpha K_{T}$~($P_{\rm th}$) has noticeable volume dependence. At fixed 
temperature, it first decreases, reaches a minimum at about $V/V_0=0.8$, 
then increases. Such behavior originates in the pressure dependence of the 
thermal expansivity~(Fig.~\ref{fig:alpha}(b)) and bulk 
moduli~(Fig.~\ref{fig:ks}(b)). This feature has also been observed in 
Ref.~\onlinecite{dorogokupets}, as
shown in Fig.~\ref{fig:thp}(b). However it is missing in the previous 
{\it ab initio} calculation.\cite{xiang}

%
%

\begin{figure}
\includegraphics[width=0.45\textwidth]{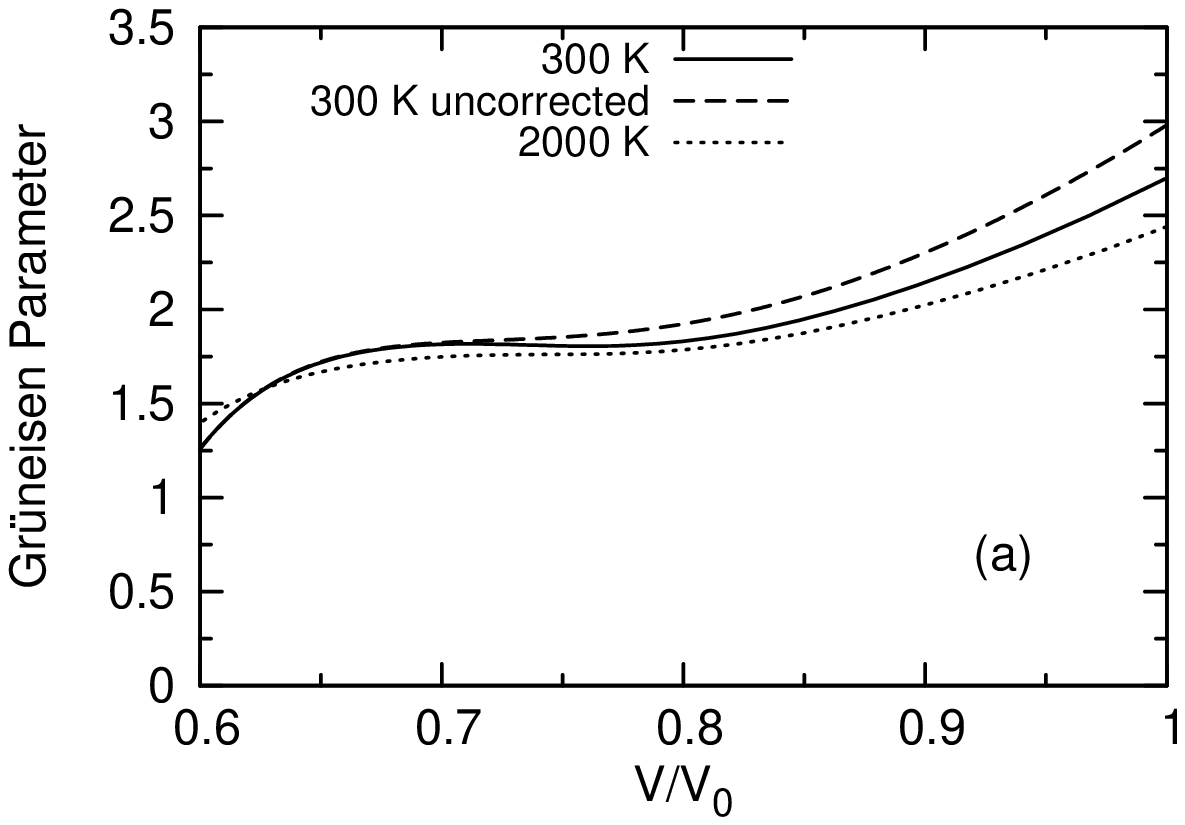}
\includegraphics[width=0.45\textwidth]{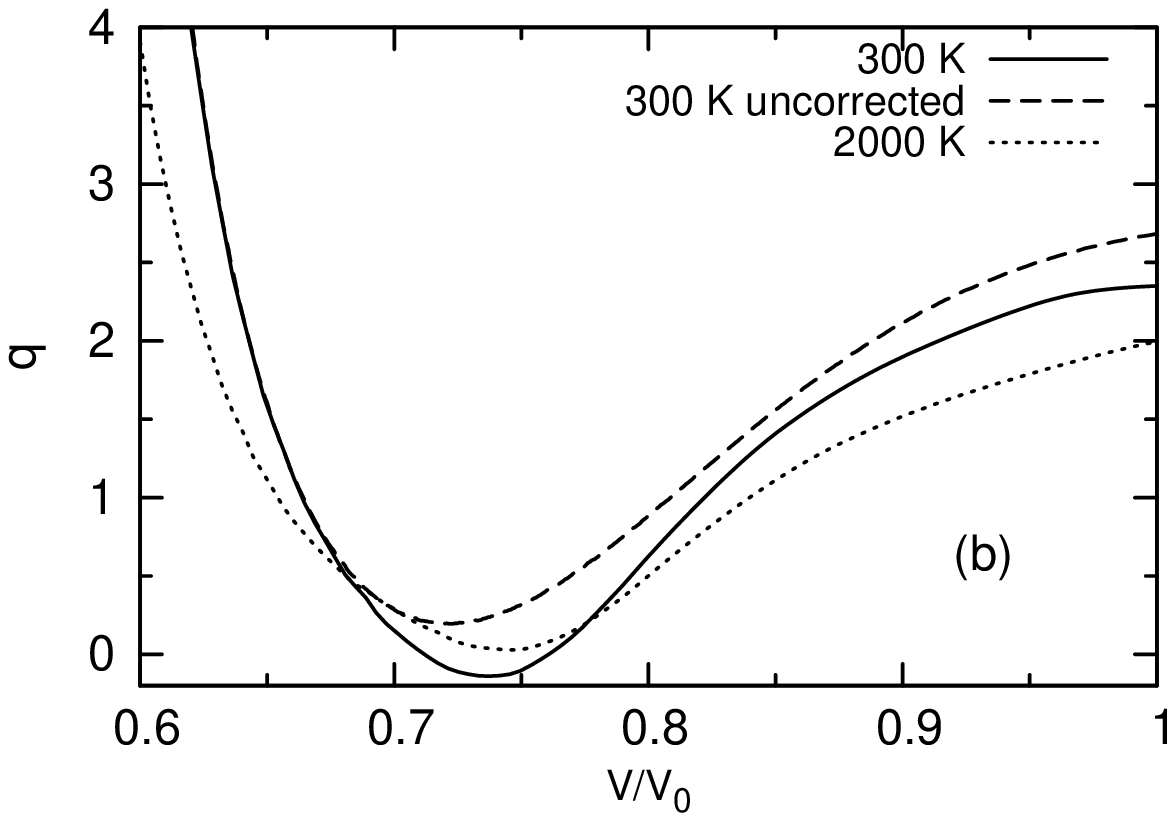}
\caption{Thermal Gr\"uneisen parameter (a) volume dependence at
fixed temperature; (b) the corresponding $q$.}
\label{fig:gruneisen}
\end{figure}

Thermal Gr\"uneisen parameter $\gamma=\frac{\alpha K_T V}{C_V}$ is an 
important quantity. Empirically it is often assumed to be independent of 
temperature. Its volume dependence is described by a parameter 
$q$=$\frac{\partial \ln \gamma}{\partial \ln V}$, and $\gamma$ can be 
represented in $q$ as
$\gamma=\gamma_{0}\left(\frac{V}{V_0}\right)^{q}.$
From Mie-Gr\"uneisen relation, it is obvious that $q$ is related to the
volume dependence of $\alpha K_T$. If $q$ equals 1, $\alpha K_T$ is 
independent of volume. If $q$ is greater than 1, $\alpha K_T$ gets smaller 
as volume decreases. In Ref.~\onlinecite{mcqueen} $q$ is assumed to equal $1$. 
Fei {\it et al.}\cite{fei2007} determined $\gamma$ by fitting the measured
$P$-$V$-$T$ data to the Mie-Gr\"{u}neisen relation up to $27$ GPa. They gave 
$\gamma_0$=$2.72$ and $q$=$0.5$. Zha {\it et al.}\cite{zha} extended 
measurements to $80$ GPa and $1900$ K. Their fit gave $\gamma_0$=$2.75$ and 
$q$=$0.25$. Our calculation indicates that the temperature dependence 
of $\gamma$ is small. The volume 
dependence of $\gamma$ is shown in Fig.~\ref{fig:gruneisen}. The 
uncorrected DFT calculation tends to overestimate $\gamma$. At ambient 
condition $\gamma_0$ equals $2.87$. 
After corrections, $\gamma_0$=$2.70$. The corresponding $q$ equals $2.35$, 
much larger than the value obtained in 
Ref.~\onlinecite{fei2007} and \onlinecite{zha}. We notice previous DFT
calculation on gold\cite{taku-au} also gives much larger $q$ than the value
in Ref.~\onlinecite{fei2007}. This is  probably due to the 
small pressure range explored in Ref.~\onlinecite{fei2007}, and limited 
number of data points measured in Ref.~\onlinecite{zha}.
%
%

\begin{figure}
\includegraphics[width=0.45\textwidth]{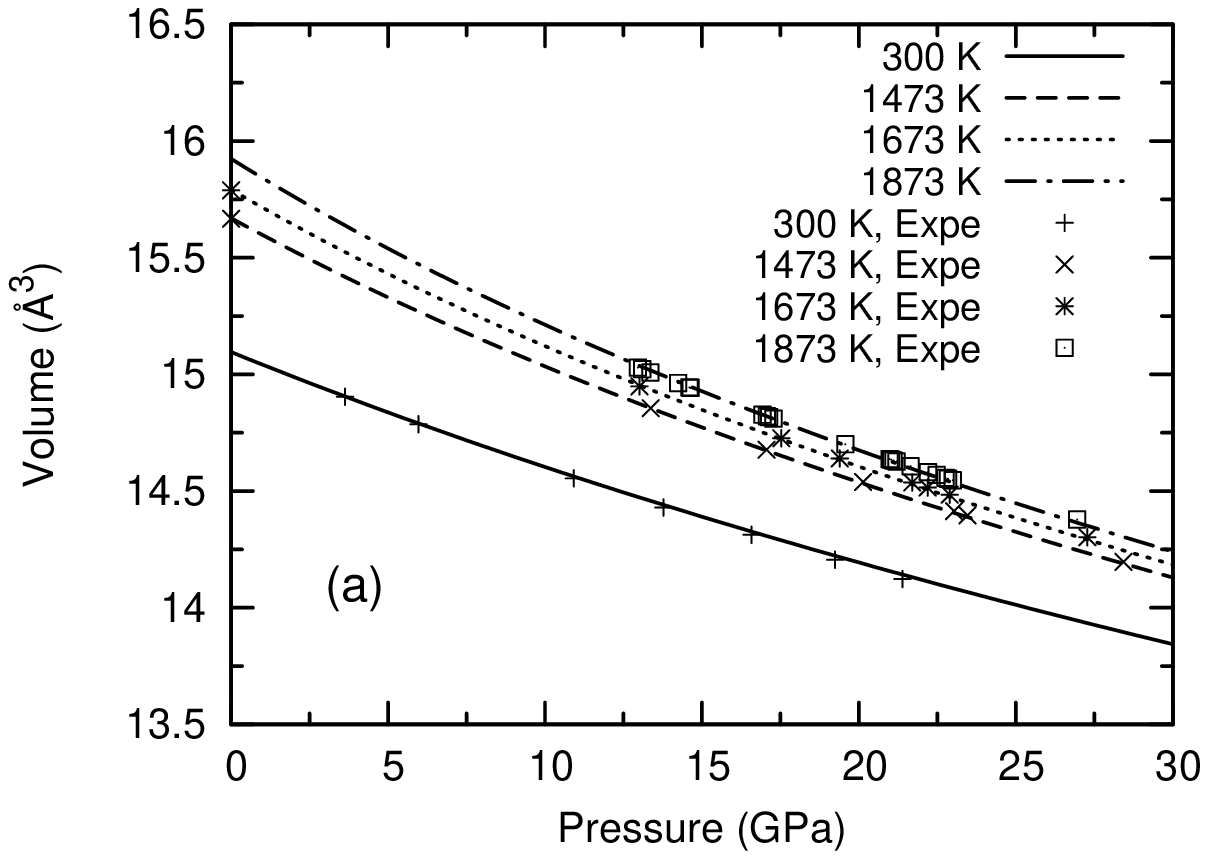}
\includegraphics[width=0.45\textwidth]{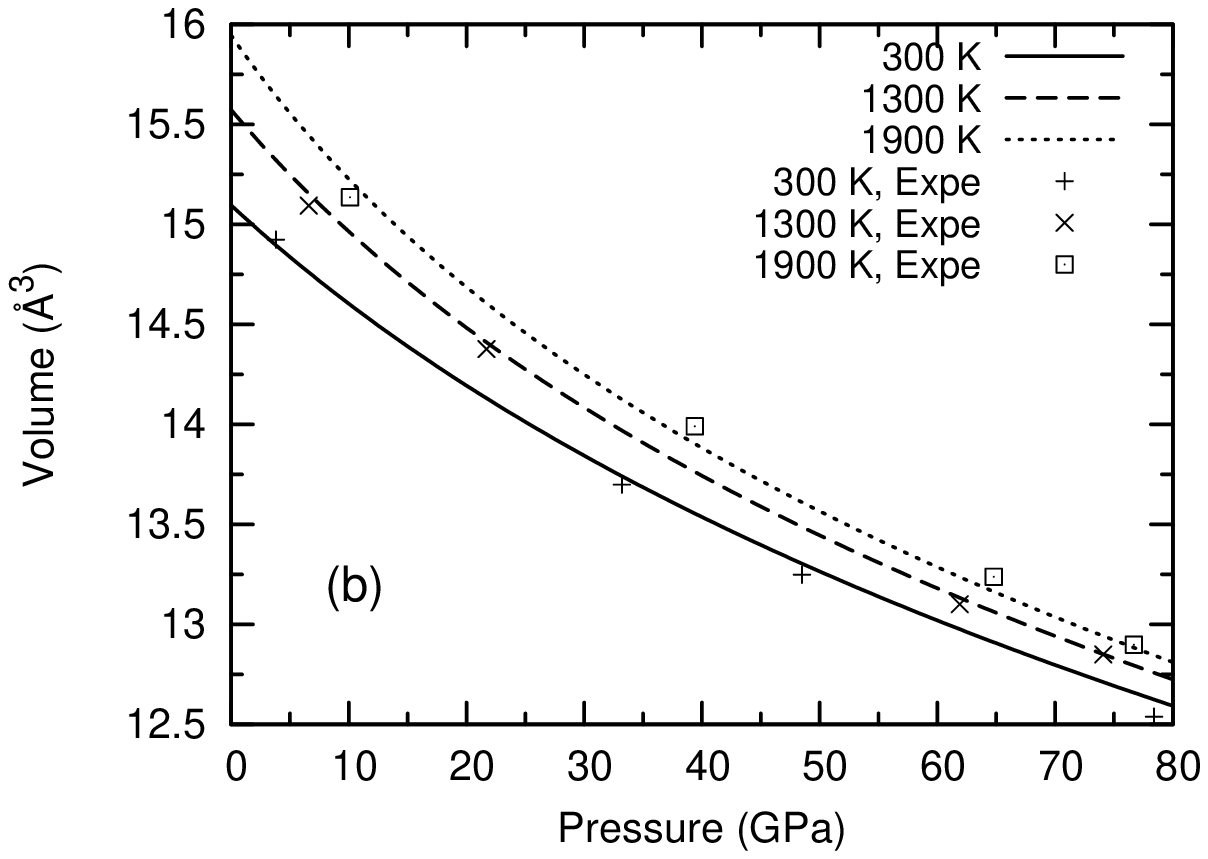}
\caption{High temperature isotherms after corrections. Lines correspond to 
the calculated isotherms. The experimental data in (a) are taken from 
Ref.~\onlinecite{fei2004}, except those at $P$=$0$ GPa, which are obtained 
by integrating the thermal expansivity listed in Ref.~\onlinecite{kirby}.
Data in (b) are from Ref.~\onlinecite{zha}.} 
\label{fig:high-t}
\end{figure}

Adding the thermal pressure to the room temperature isotherm, we get the
thermal EOS of platinum, as shown in Table~\ref{tab:isochor}. We compare
our results with two DAC measurements in Fig.~\ref{fig:high-t}. Within the
error of the experiments, the agreement is reasonably good. For
convenience of interpolation, parametric forms of the thermal EOS are listed 
in Table~\ref{tab:para}.

%
%
\begin{table}
\caption{Pressure~(in GPa) as a function of compression~(1-$V/V_0$, $V_0$ 
is the experimental volume at ambient condition. $15.095$ \AA$^3$) and 
temperature (in K), deduced from the corrected Gibbs free energy. Pressures
above melting point is not shown. The melting point is determined using 
$T_m(P)=2057+27.2\times P-0.1497\times P^2$ K up to 70 GPa from
Ref.~\onlinecite{kavner}}
\begin{ruledtabular}
\begin{tabular}{c|ccccccccccc}
$1-V/V_0$ & $300$ & $500$ & $1000$ & $1500$ & $2000$ & $2500$ 
& $3000$ & $3500$ & $4000$ & $4500$ & $5000$ \\
\hline
0.00 & 0.00  & 1.51  & 5.31  & 9.14  & 12.97 \\
0.05 & 16.22 & 17.62 & 21.20 & 24.83 & 28.48 & 32.16 \\
0.10 & 38.32 & 39.64 & 43.05 & 46.52 & 50.03 & 53.57 & 57.14 \\
0.15 & 68.41 & 69.67 & 72.97 & 76.34 & 79.77 & 83.24 & 86.73 & 90.25 \\ 
0.20 & 109.46 & 110.71 & 113.99 & 117.37 & 120.82 & 124.31 & 127.83 
& 131.39 & 134.98 \\
0.25 & 166.45 & 167.74 & 171.15 & 174.69 & 178.29 & 181.94 & 185.63 & 189.36 
& 193.13 & 196.94 & 200.81 \\
0.30 & 247.37 & 248.72 & 252.34 & 256.07 & 259.88 & 263.73 & 267.64 & 271.58 
& 275.57 & 279.58 & 283.63 \\
0.35 & 362.30 & 363.65 & 367.31 & 371.10 & 374.98 & 378.91 & 382.90 & 386.94 
& 391.01 & 395.12 & 399.25 \\
0.40 & 525.86 & 526.93 & 530.04 & 533.39 & 536.86 & 540.40 & 543.99 & 547.62  
& 551.28 & 554.99 & 558.76 \\
\end{tabular}
\end{ruledtabular}
\label{tab:isochor}
\end{table}
%
%
%

%
\begin{table}
\caption{Parametric form of the thermal EOS,  
$P(V)$=$P_0+\frac{3}{2} K_0 
[(V/V_0)^{-\frac{7}{3}}-(V/V_0)^{-\frac{5}{3}}]
\cdot\left\{
1+\frac{3}{4}(K'_0-4)[(V/V_0)^{-\frac{2}{3}}-1]+\frac{3}{8}
[K_0 K''_0+(K'_0-3)(K'_0-4)+\frac{35}{9}]
[(V/V_0)^{-\frac{2}{3}}-1]^2
\right\}$. At high temperatures, the equilibrium volume may exceed the 
largest volume we calculate. For better accuracy we fit the $P$-$V$-$T$ 
data in three different pressure-temperature intervals: (1) 
$0$-$100$~GPa and $0$-$2000$ K, (2) $50$-$250$~GPa and $0$-$3000$ K, 
(3) $150$-$550$~GPa and $0$-$5000$ K. 
$P_0$ denotes the starting pressure of the corresponding interval. $V_0$,
$K_0$, $K'_0$, and $K''_0$ are temperature dependent parameters, and 
are fitted to a 4th order polynomial $a_0+a_1 t +a_2 t^2+a_3 t^3 +a_4 t^4$, 
where $t$=$T/1000$. They have a formal correspondence to the usual 4th order 
BM EOS parameters, which are defined at $P_0=0$ GPa.}
\begin{ruledtabular}
\begin{tabular}{cccccc}
(1) & $a_0$ & $a_1$ & $a_2$ & $a_3$ & $a_4$ \\
\hline
$V_0$(\AA$^3$) & 14.9924 & 0.295837 & 0.194441 & -0.0917141 & 0.024365 \\
$K_0$(GPa) & 290.539 & -45.4082 & -9.38792 & 5.09573     & -1.40266  \\
$K'_0$ & 5.11956 & 0.52903 & 0.0733263 & -0.0195011 & 0.0229666 \\
$K''_0$(GPa$^{-1}$)&-0.0275729&-0.0120014&-0.0114928&0.00672243&-0.00359317\\
\hline
(2) & $a_0$ & $a_1$ & $a_2$ & $a_3$ & $a_4$ \\
\hline
$V_0$(\AA$^3$) & 13.2246 & 0.128227 & 0.049052 & -0.0160359 & 0.00241857 \\
$K_0$(GPa) & 523.48 & -30.3849 & -3.86087  & 1.31313 & -0.222027  \\
$K'_0$ & 4.24183 & 0.217262 & -0.0235333 & 0.00944835 & -0.000371746 \\
$K''_0$(GPa$^{-1}$)& -0.00125873 & -0.00268918&  2.13874e-05& -3.57657e-05& 
-1.75847e-05\\
\hline
(3) & $a_0$ & $a_1$ & $a_2$ & $a_3$ & $a_4$ \\
\hline
$V_0$(\AA$^3$) & 11.4929& 0.0672156 & 0.0119585 & -0.00243269 & 0.000219022 \\
$K_0$(GPa) & 951.004 & -21.0874 & -2.84254 & 0.654708 & -0.0639296 \\
$K'_0$ & 4.31383 & 0.05775 & -0.00505386 & 0.00245414 & -0.000167453 \\
$K''_0$(GPa$^{-1}$)& -0.00588145& -0.00130468& 0.000221904 & -6.51359e-05& 4.99978e-06\\
\end{tabular}
\end{ruledtabular}
\label{tab:para}
\end{table}

\begin{figure}
\includegraphics[width=0.45\textwidth]{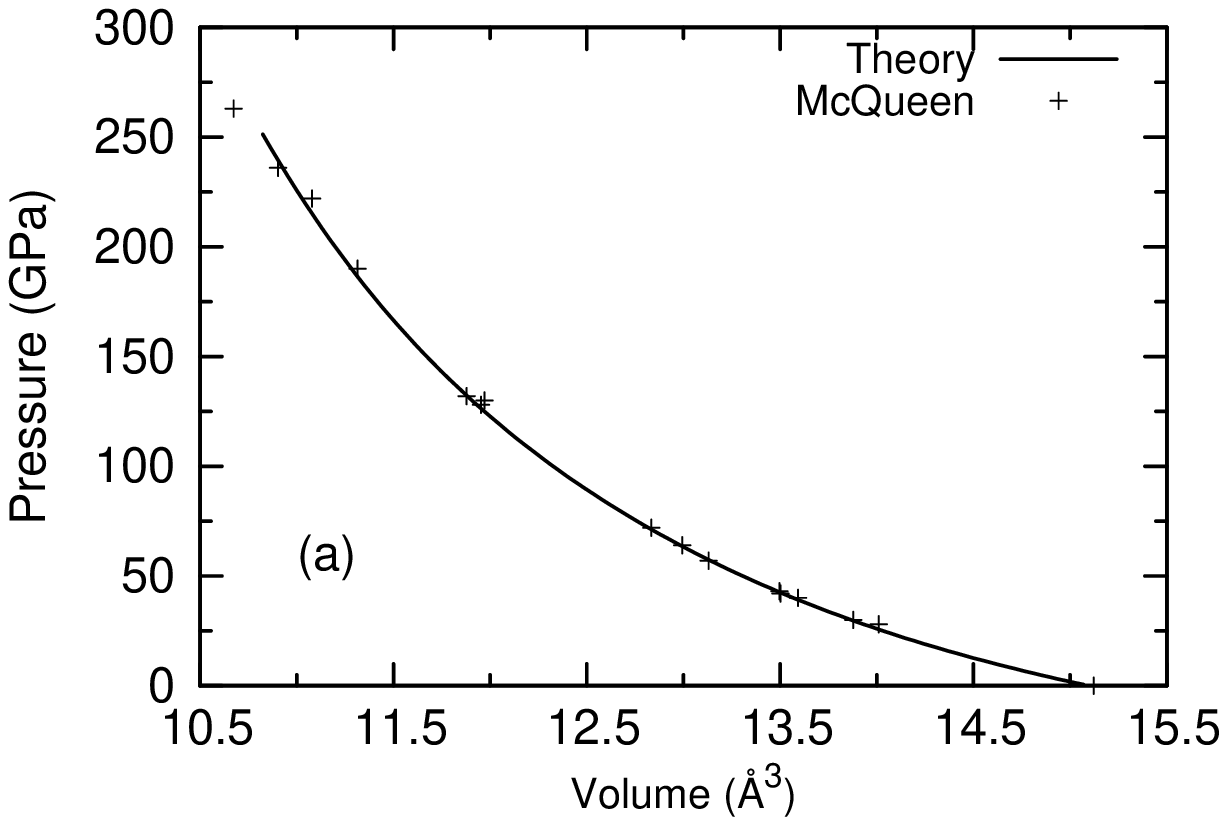}
\includegraphics[width=0.45\textwidth]{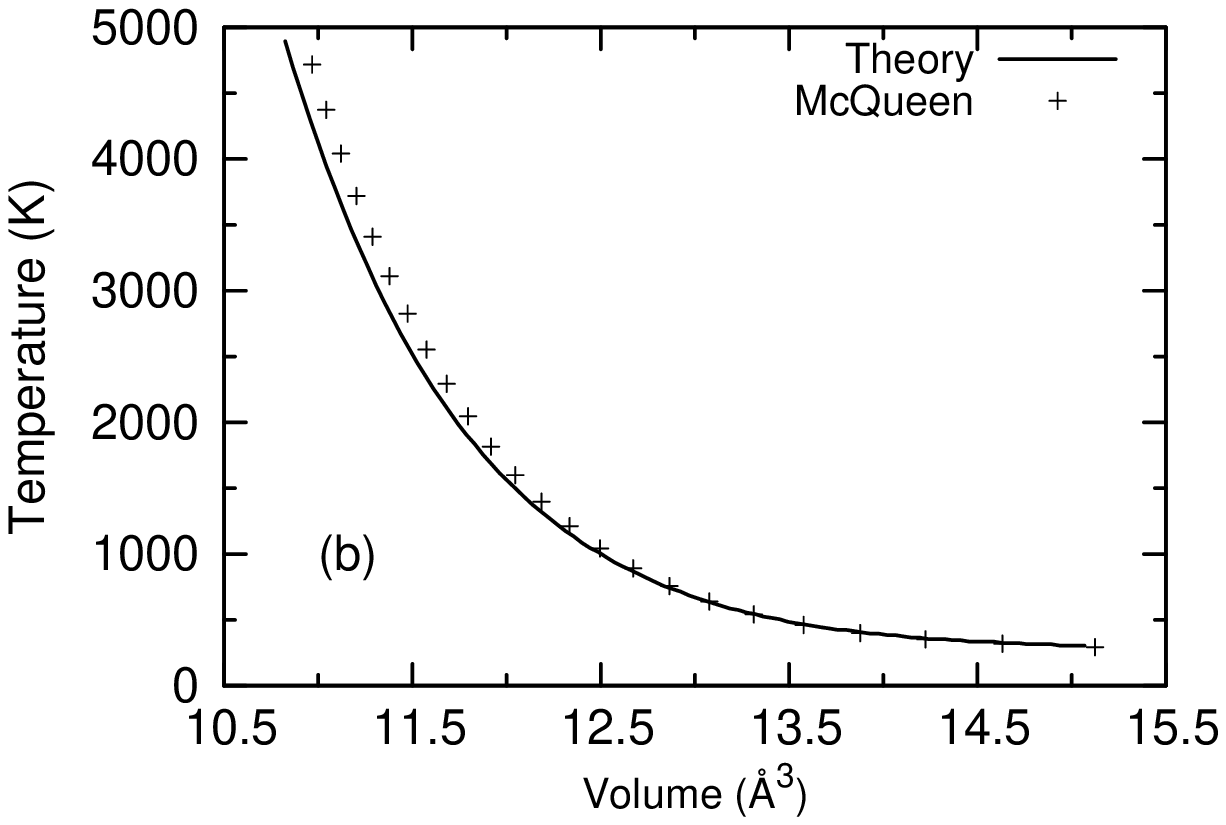}
\caption{(a) Theoretical shock Hugoniot compared with the expermental data
from McQueen {\it et al.}~(Ref.~\onlinecite{mcqueen}). (b) Temperature along 
the Hugoniot.}
\label{fig:hugoniot}
\end{figure}

The $P$-$V$-$T$ thermal EOS we obtained are very similar to the one in
Ref.~\onlinecite{dorogokupets} below $100$ GPa, This is expectable as we
used the $300$ K isotherm in Ref.~\onlinecite{dorogokupets} as the reference
to correct the room temperature Gibbs energy, and the thermal properties 
calculated from both approaches agree well with the experiments. In this 
$P$-$T$ range, the uncertainty of our EOS is comparable to the one in 
Ref.~\onlinecite{dorogokupets} . Above $100$ GPa, the uncertainty is 
about $1.4$~\%, which is difference between the LAPW~(LDA) and pseudo-1 
static EOS. Other sources of error, e.g. convergence uncertainty~($0.5$~\%)
and ignoring spin-orbit effect~($0.7$~\%) are smaller effects.
To check the accuracy of our thermal EOS at high 
pressures, we start from the corrected Gibbs energy and compute the 
theoretical Hugoniot by solving the Rankine-Hugoniot equation:
\begin{equation}
E_{H}(V,T)-E_i(V_0,T_i)=(P_{H}(V,T)+P_0(V_0,T_i))\frac{V_0-V}{2},
\label{eq:hugoniot}
\end{equation}
where $E_{H}(V,T)$, $P_{H}(V,T)$ are the internal energy, pressure
at volume $V$ and temperature $T$. $E_{i}(V_0,T_i)$, $P_0(V_0,T_i)$
are the internal energy, pressure at the initial volume $V_0$ and
temperature $T_i$. The results are shown in Fig.~\ref{fig:hugoniot}.
The predicted Hugoniot pressure is in good agreement with measurements.
The temperature predicted by DFT is lower than the empirically deduced value 
in Ref.~\onlinecite{mcqueen}. The reduction in Ref.~\onlinecite{mcqueen} 
neglects the electronic thermal pressure, and this 
may cause overestimating Hugoniot temperature.\cite{holmes}

We end this section by comparing our room temperature isotherm with that 
of Holmes {\it et al.}. Below 70 GPa, they are almost identical. At high
pressures~($200$-$550$ GPa), the pressure from our EOS is about $3$~\% lower 
than the one from Holmes {\it et al.}. Holmes {\it et al.} used LMTO with 
the atomic-sphere approximation to get the static EOS. In principle, the full 
potential LAPW method used in this study is more accurate. It seems the EOS 
of Holmes {\it et al.} overestimates pressure systematically at high 
compression ratio. But the magnitude is much smaller than the pressure offset 
needed to compensate the discrepancy between Mao {\it et al.}'s experiment 
and seismological extrapolation. The real cause of the discrepancy might be a 
combination of several factors.
%
%

%
%
\section{Conclusions}
In this paper, we report our calculations on the static and thermal EOS 
of platinum using DFT with different exchange correlation functionals. 
Contrary to previous reports, we find the room temperature isotherm 
computed with LDA lies below, and nearly parallel to the experimental 
compression data. We study the lattice dynamics of platinum within QHA, and 
find the electronic temperature dependence of vibrations plays a noticeable 
role in determining the thermal properties of platinum. Combining the
experimental data with DFT calculations, we propose a consistent thermal
EOS of platinum, up to $550$ GPa and $5000$ K, which can be used 
as a reference for pressure calibration.
%
\acknowledgments
We thank P. B. Allen, P. I. Dorogokupets, A. Floris, B. B. Karki for 
discussions and help; A. Dewaele for suggestions and taking 
Ref.~\onlinecite{dor2} to our attention; Y. W. Fei for sending us 
Ref.~\onlinecite{zha}. We are in debt to the anonymous referees for careful 
reviews. The pseudopotential calculations were performed at the Minnesota 
Supercomputing Institute (MSI) with the Quantum ESPRESSO 
package~(http://www.pwscf.org). The LAPW calculations were performed at
Brookhaven National Laboratory (BNL) with the WIEN2k 
package~(http://www.wien2k.at). TS was supported by NSF ITR Grant No.
ATM0426757. RMW, KU, and ZW were supported by 
NSF/EAR 0230319, 0635990, and NSF/ITR 0428774 (VLab).
\bibliography{citation}

\begin{thebibliography}{48}
\bibitem[{ber()}]{bercegeay-note}
\bibinfo{note}{This TM pseudopotential is constructed from $5d^{9}
  6s^{0.95}6p^{0.05}$ atomic configuration, the cutoff radius of $6s$ is $2.6$
  a.u., which is probably too large. The HGH pseudopotential is constructed
  from $5s^{2}5p^{6}5d^{10}$ and does not have this problem.}
\end{thebibliography}
\end{document}